\documentclass[aps,pra,reprint,superscriptaddress]{revtex4-2}
\usepackage{natbib}

\usepackage{color}
\usepackage{amsmath}
\usepackage{amssymb}
\usepackage{bbm}
\usepackage{framed}
\definecolor{shadecolor}{rgb}{0.85, .85, 0.85}
\usepackage{graphicx}
\usepackage{multirow}
\usepackage[utf8]{inputenc}
\usepackage[T1]{fontenc}
\usepackage{CJKutf8}

\definecolor{darkgreen}{rgb}{0.05, .65, 0.05}

\def\dlangle{\langle\!\langle}
\def\drangle{\rangle\!\rangle}
\def\bigdlangle{{\big\langle\kern-.23em\big\langle}}
\def\bigdrangle{{\big\rangle\kern-.23em\big\rangle}}
\def\Bigdlangle{{\Big\langle\kern-.35em\Big\langle}}
\def\Bigdrangle{{\Big\rangle\kern-.35em\Big\rangle}}
\def\biggdlangle{{\bigg\langle\kern-.35em\bigg\langle}}
\def\biggdrangle{{\bigg\rangle\kern-.35em\bigg\rangle}}
\def\Biggdlangle{{\Bigg\langle\kern-.35em\Bigg\langle}}
\def\Biggdrangle{{\Bigg\rangle\kern-.35em\Bigg\rangle}}

\def\biggdexpct#1{\biggdlangle{#1}\biggdrangle}

\def\intzinf{\int_{0}^\infty\!\!}

\def\nablaR#1{\nabla_{\vect{R}_{#1}}}

\def\linklangle{{\langle\kern-.28em\scalebox{0.35}[1]{$-$}\kern-.18em\langle}}
\def\linkrangle{{\rangle\kern-.18em\scalebox{0.35}[1]{$-$}\kern-.28em\rangle}}
\def\biglinklangle{{\kern.23em\big\langle\kern-.59em\scalebox{0.35}[1]{$-$}\kern-.38em\big\langle\kern.23em}}
\def\biglinkrangle{{\kern.23em\big\rangle\kern-.38em\scalebox{0.35}[1]{$-$}\kern-.59em\big\rangle\kern.23em}}
\def\Biglinklangle{{\kern.0em\Big\langle\kern-.49em\scalebox{0.38}[1]{$-$}\kern-.18em\Big\langle\kern.0em}}
\def\Biglinkrangle{{\kern.0em\Big\rangle\kern-.18em\scalebox{0.38}[1]{$-$}\kern-.49em\Big\rangle\kern.0em}}
\def\bigglinklangle{{\kern.0em\bigg\langle\kern-.62em\scalebox{0.54}[1]{$-$}\kern-.18em\bigg\langle\kern.0em}}
\def\bigglinkrangle{{\kern.0em\bigg\rangle\kern-.18em\scalebox{0.54}[1]{$-$}\kern-.62em\bigg\rangle\kern.0em}}
\def\Bigglinklangle{{\kern.0em\Bigg\langle\kern-.66em\scalebox{0.66}[1]{$-$}\kern-.22em\Bigg\langle\kern.0em}}
\def\Bigglinkrangle{{\kern.0em\Bigg\rangle\kern-.22em\scalebox{0.66}[1]{$-$}\kern-.66em\Bigg\rangle\kern.0em}}

\def\I[#1]#2{\mathbbm{1}^{(#1)}_{#2}}

\def\vect#1{\ensuremath{\mathbf{#1}}}

\def\sgn{\ensuremath{\mathrm{sgn}}}

\def\cT{\ensuremath{\mathcal{T}}}

\def\x0{\ensuremath{\mathbf{x}_\mathrm{0}}}
\def\epsr{\ensuremath{\epsilon_{\mathrm{r}}}}

\def\subEM{_\mathrm{\scriptscriptstyle EM}}
\def\subTE{_\mathrm{\scriptscriptstyle TE}}

\def\subCP{_\mathrm{\scriptscriptstyle CP}}

\def\supTE{^\mathrm{\scriptscriptstyle (TE)}}

\def\etal{\textit{et.~al}}

\def\expct#1{\!\left\langle{#1}\right\rangle}

\definecolor{darkblue}{rgb}{0.0,0.0,0.5}
\usepackage[colorlinks,citecolor=darkblue,linkcolor=darkblue]{hyperref}

\begin{document}

\author{Jonathan B. Mackrory}
\affiliation{Department of Physics and Oregon Center for Optical, Molecular and Quantum Science, 
  1274 University of Oregon, Eugene, Oregon 97403-1274}
\author{He Zheng}
\affiliation{Department of Physics and Oregon Center for Optical, Molecular and Quantum Science, 
  1274 University of Oregon, Eugene, Oregon 97403-1274}
\author{Daniel A. Steck}
\affiliation{Department of Physics and Oregon Center for Optical, Molecular and Quantum Science, 
  1274 University of Oregon, Eugene, Oregon 97403-1274}

\title{Pathwise Differentiation of Worldline Path Integrals}
\date{\today}

\begin{abstract}
  The worldline method is a powerful numerical path-integral
  framework for computing Casimir
  and Casimir--Polder energies.
  An important challenge arises when one desires derivatives
  of path-integral quantities---standard finite-difference techniques,
  for example,
  yield results of poor accuracy.
  In this work we present methods for computing derivatives
  of worldline-type path integrals of scalar fields
  to calculate forces, energy curvatures, and torques.  
  In Casimir--Polder-type path integrals, which require 
  derivatives with respect to the source point of the paths, 
  the derivatives can be computed 
  by a simple reweighting of the path integral.  
  However, a partial-averaging technique is necessary
  to render the differentiated path integral computationally efficient.
  We also discuss the computation of Casimir forces, curvatures, 
  and torques between macroscopic bodies.
  Here a different method is used, involving summing over the derivatives of
  all the intersections with a body; again, a different partial-averaging
  method makes the path integral efficient.
  To demonstrate the efficiency of the techniques, we give the results
  of numerical implementations of these worldline methods in
  atom--plane and plane--plane geometries.
  Being quite general, the methods here should apply to path integrals
  outside the worldline context (e.g., financial mathematics).

\end{abstract}

\maketitle

\section{Introduction}

The Casimir effect is a fundamental consequence of quantum fields interacting with boundaries~\cite{Casimir1948a}.
It typically arises as an attractive force between material bodies
due to modifications of vacuum fluctuations 
by the presence of the bodies~\cite{CasimirPolder1948,Lifshitz1956}.  
An important experimental technique for measuring Casimir-type forces monitors
the change in the frequency of a mechanical oscillator due to the presence of 
another dielectric body: such experiments are sensitive to the \textit{curvature}
of the Casimir energy.  
This technique has been applied, for example, to  
Casimir-force measurements in microelectromechanical systems~\cite{Chan2001b}, and 
also to Casimir--Polder measurements with Bose-Einstein
condensates in a magnetic trap near a dielectric mirror~\cite{Harber2005}.  
Theoretical predictions of Casimir forces (and corresponding frequency shifts in experiments)
require the computation of gradients of 
the Casimir energy. In general geometries, this requires a numerical approach.

The worldline path-integral method in particular
is a promising framework for the numerical computation of Casimir energies~\cite{Gies2003,Gies2006a,Gies2006b,Gies2006c,Gies2006d,Klingmuller2008,Weber2009,Fosco2010,Weber2010,Aehlig2011,Schafer2012,Mazur2015,Schafer2016,Mackrory2016}.
It continues to be a useful method in various areas of 
quantum field theory~\cite{Schneider18,FranchinoVinas19,Corradini20,Rajeev21,Bastianelli24}.
In this method one generates an ensemble of closed stochastic paths and evaluates
the Casimir energy as a functional average over the material properties along the paths.
The worldline method has been thus far primarily applied in the case of scalar fields interacting with 
Dirichlet boundary conditions~\cite{Gies2006a,Gies2006b,Gies2006c,Gies2006d,Klingmuller2008}.  
For example, this work has been applied to computing Casimir forces and torques
for cylinder-plate, sphere-plate, and inclined-plate geometries~\cite{Weber2009,Weber2010},
as well as to computing the stress-energy tensor for parallel planes~\cite{Schafer2012,Schafer2016}.
More recently, the worldline method has been extended to incorporate the coupling of
the electromagnetic field to magnetodielectric media~\cite{Mackrory2016}.
In that reference, the electromagnetic worldline path integrals are only exact
in symmetric geometries where the 
vector electromagnetic field splits into two uncoupled scalar fields.

In connecting the worldline method to experiments, where one is interested
in forces and curvatures as numerical quantities,
it is therefore important to consider the numerical differentiation of worldline
path integrals.
Additionally, the worldline path integral 
for the part of the Casimir--Polder potential 
corresponding to the transverse-magnetic
(TM) polarization involves the second derivative of a worldline path integral~\cite{Mackrory2016}.
Similarly, the calculation of the stress-energy tensor via worldline methods
involves derivatives of worldline path integrals~\cite{Scardiccio2006,Schafer2012,Schafer2016}.  
Due to the importance of
estimating gradients of Monte Carlo path averages 
in other areas such as mathematical finance,
there has already been a substantial effort towards developing
general methods for differentiating path integrals~\cite{Glasserman2004,Asmussen2007}.   
Since the methods in this paper apply quite generally, they may
be applicable to such problems.

The most straightforward way to compute a numerical derivative
is the finite-difference method.  
While finite-difference approximations are quite general and easy to implement,
they have poor convergence properties for gradients of stochastic quantities~\cite{Glasserman2004,Asmussen2007}. 
In computing a first derivative, for example, one needs to compare functional values
of two slightly different paths, and then divide by a small parameter that characterizes the
difference between them.  In such averages over stochastic paths, the difference
in the contributions of these paths may be large---in a worldline path integral, for example,
one path may intersect a dielectric while its slightly shifted counterpart misses it, but these
differences are precisely what determine the Casimir force. 
Thus, the finite-difference method can lead to a path average with a large sample variance,
requiring extensive 
averaging---and thus excessive computing resources---to arrive 
at an accurate result. The problem is even worse for successive derivatives.

In this paper we present numerically efficient methods for computing gradients 
of worldline path integrals in two classes of problems.
First, we consider derivatives of Casimir--Polder path integrals, where the derivatives
are taken with respect to the source point of the paths.
In this case, a reweighted path integral serves to compute the derivatives,
but a partial-averaging technique is needed to avoid convergence problems
in the limit of continuous-time paths.
This method applies rather generally to numerical path integrals differentiated
with respect to the source point. For example, it applies whenever
the path functional is independent of the motion of the path in some
neighborhood of the source point.  It also applies if the path functional
may be approximately averaged over some time interval around the source point.

In the second class of problems, we consider the calculation 
of Casimir forces, curvatures, and torques due to extended material bodies.  
A particularly relevant method is that of Weber and
Gies, who computed
forces in the sphere--plane and cylinder--plane geometries 
with Dirichlet boundary
conditions~\cite{Weber2010}.  
In that work paths are shifted until they graze the planar surface, 
taking advantage of the freedom of the integral over $\vect{x}_0$.  
The force on the planar surface is then computed by integrating 
over the times when the path intersects the spherical or cylindrical surface.
By contrast, we consider a generally valid method that sums over the
derivative of an estimate
of the path average or value between two successive points in the discrete path.
Again, a partial-averaging method accelerates convergence.

\section{Derivatives of Worldline Casimir--Polder Energies}

To serve as a concrete example of path-integral differentiation, 
we will consider Casimir-type energies for the scalar field
corresponding to the transverse-electric (TE) polarization of the electromagnetic field
coupled to a dielectric~\cite{Mackrory2016}.   
In the strong-coupling (perfect-conductor) limit, this path integral imposes the same
Dirichlet boundary conditions as in the path integrals considered by Gies~\etal~\cite{Gies2003}.  

The basic expression for the Casimir energy of this scalar field in 
a dielectric-permittivity background $\epsilon(\mathbf{r})$ is
(without renormalization)
\begin{align}
  E &= -\frac{\hbar c}{2(2\pi)^{D/2}}\int_{0}^{\infty}
  \frac{d\cT}{\cT^{1+D/2}}\int_{\text{all space}} \!\!\!\!\!d\vect{x}_0 \nonumber\\
  &\hspace{2.5cm}\times
  \biggdlangle \frac{1}{\langle \epsr(\vect{x})\rangle^{1/2}}
  \biggdrangle_{\vect{x}(t)} ,
  \label{eq:casimir_energy}
\end{align}
or equivalently,
\begin{align}
  E &= -\frac{\hbar}{(2\pi)^{(D+1)/2}}\int_0^\infty\! ds\int_{0}^{\infty}
  \frac{d\cT}{\cT^{(D+1)/2}}\int_{\text{all space}} \!\!\!\!\!d\vect{x}_0 \nonumber\\
  &\hspace{2.5cm}\times
  \biggdlangle e^{-s^2\cT \langle \epsr(\mathbf{x})\rangle/2c^2}\biggdrangle_{\mathbf{x}(t)} .
  \label{eq:casimir_energy_exp}
\end{align}
In these expressions, $D$ is the number of spacetime dimensions,
but the temporal dimension has been explicitly removed at this stage.
This expression also
assumes a dispersion-free dielectric at zero temperature, but it
(and the following treatment) are
readily generalized to incorporate dispersion and temperature effects \cite{Mackrory2016}.
In this expression, the double angle brackets $\dlangle \cdots\drangle$ denote
an ensemble average over closed Brownian paths $\vect{x}(t)$
(in $D-1$ spatial dimensions) that
begin and end at $\vect{x}(0)=\vect{x}(\cT)=\vect{x}_0$.  
The single-angle-bracket notation $\,\expct{\epsr(\vect{x})}$ indicates an
average of the relative dielectric permittivity $\epsr:=\epsilon/\epsilon_0$
along the path $\vect{x}(t)$:  
\begin{equation}
  \langle \epsr \rangle := \frac{1}{\cT}\int_0^\cT \!\!dt\, 
  \epsr[\vect{x}(t)].
  \label{eq:path_average}
\end{equation}
The path increments $d\vect{x}(t)$
are Gaussian, with moments $\dlangle d\vect{x}(t)\drangle = 0$ and
$\dlangle dx_i(t)\,dx_j(t)\drangle= \delta_{ij}\,dt$.  
The energy in these expressions must still be renormalized by 
subtracting the zero-body energy 
(obtained by replacing $\mathbf{x}$ by $\mathbf{x}_0$ in the stochastic average)
and the one-body energy
(obtained as a limit of widely separated bodies),
removing the divergence at $\cT=0$.  
Physically, the zero-body term represents the energy associated with
a uniform dielectric background of permittivity $\epsilon(\vect{x}_0)$.

The Casimir--Polder potential for an atom interacting with a dielectric body
arises by treating the atom as a localized 
perturbation to the dielectric,
$\epsr(\vect{x})\longrightarrow \epsr(\vect{x}) + \alpha_0\delta(\vect{x}-\x0)/\epsilon_0$,
and keeping the energy change to linear order 
in the (static) atomic polarizability $\alpha_0$.
Again, this prescription ignores dispersion and temperature, but
such effects are straightforward to include \cite{Mackrory2016}.
The resulting path integral following from (\ref{eq:casimir_energy}) is
\begin{align}
  V\subCP(\x0) =&\, \frac{\hbar c\alpha_0}{4(2\pi)^{D/2}\epsilon_0}\int\frac{d\cT}{\cT^{1+D/2}}
  \biggdlangle \frac{1}{\langle \epsr\rangle^{3/2}}\biggdrangle_{\vect{x}(t)}\!\!\!.
  \label{eq:CP_energy}
\end{align}
Again, this expression must be renormalized by subtracting the zero-body energy,
which amounts to subtracting 1, 
assuming that the atom is not embedded in a dielectric medium (i.e., $\epsr=1$ in a neighborhood
of $\vect{x}_0$).
This expression for the Casimir--Polder potential is similar
to that of the Casimir energy (\ref{eq:casimir_energy}), 
but the potential here is solely a function 
of paths emanating from the atomic position $\vect{x}_0$.

\subsection{Finite Differences}

In path integrals such as Eq.~(\ref{eq:CP_energy}) for
the Casimir--Polder potential, one can compute the Casimir--Polder force 
$\mathbf{F}\subCP=-\nabla E\subCP$
by differentiating with respect to the source point $\vect{x}_0$ of the paths.  
A simple, direct way to compute such a gradient numerically is  
the finite-difference method.  
In this technique the derivative is computed in terms of the contribution of
each path
$\vect{x}(t)$ and its shifted counterparts of the form 
$\vect{x}(t)+\hat{r}_i\delta$, where $\delta$ is a small number, and $\hat{r}_i$ form
an orthonormal basis for the configuration space.  
The simplest finite difference for the first derivative, for example,
yields the following path integral for the force, from Eq.~(\ref{eq:CP_energy}):
\begin{align}
  \mathbf{F}(\x0) =& 
  -\frac{\hbar c\alpha_0}{4(2\pi)^{D/2}\epsilon_0}
  \sum_i\hat{r}_i
  \int_0^\infty\frac{d\cT}{\cT^{1+D/2}}\nonumber\\
  &\hspace{0.cm} \times
  \biggdlangle \frac{\langle \epsr[\vect{x}(t)+\hat{r}_i\delta]\rangle^{-3/2}-\langle \epsr[\vect{x}(t)]\rangle^{-3/2}}{\delta}\biggdrangle_{\vect{x}(t)}\!\!.    
\end{align}
Unfortunately, as a numerical method this path integral performs poorly.
Specifically, 
in the case of interfaces between regions of otherwise uniform dielectric permittivity,
the contribution to the force path integral 
for a given path $\vect{x}(t)$ vanishes
except when $\smash{\expct{\epsr}}$ differs for the shifted and original paths
$\vect{x}(t)+\hat{r}_i\delta$ and $\vect{x}(t)$, respectively.
Depending on how $\smash{\expct{\epsr}}$ is computed numerically, 
this difference may vary substantially among the possible paths
$\vect{x}(t)$ in the limit of small $\delta$.
For example, in the limit of an interface between vacuum and a perfect conductor,
the contribution for a given path $\vect{x}(t)$
vanishes except in the (unlikely) case that
$\vect{x}(t)+\hat{r}_i\delta$ crosses the interface but $\vect{x}(t)$ does not.
For small $\delta$, only a small subset of the paths make a nonvanishing 
contribution to the path integral, and the sample variance diverges as $\delta\longrightarrow 0$.  
This problem becomes progressively
worse when computing additional derivatives. Thus, it is important
to consider alternative methods for computing derivatives.

Because of the presence of the $\cT$ integral in the path integrals
(\ref{eq:casimir_energy}) and (\ref{eq:CP_energy}),
it is possible in particular geometries to implement finite differences
in a way that avoids the convergence problems noted above.
This follows from noting that, for example, in the case of a planar dielectric interface,
a shift in the source point of a path is equivalent to a change in the path's
running time $\cT$, as far as computing \,$\smash{\expct{\epsr}}$ is concerned.
Thus, the finite difference can effectively act on the $\cT^{-1-D/2}$ factor
instead of the path average.  
However, the rest of this paper 
will focus on methods that
apply in general geometries and to path averages
that lack the $\cT$ integral.

\subsection{Partial Averages in the Vicinity of the Source Point}
\label{sec:partial_average}
To handle derivatives with respect to the atomic position 
in the Casimir--Polder energy~(\ref{eq:CP_energy}), consider the general, schematic
form of the path integral
\begin{align}
  I &= \bigdlangle \Phi[\vect{x}(t)]\bigdrangle_{\vect{x}(t)}\nonumber\\
  &= \int \prod_{n=1}^{N-1}d\vect{x}_n \,P[\vect{x}(t)]\,\Phi[\vect{x}(t)]
  \label{eq:general_functional}
\end{align}
in terms of an arbitrary path functional $\Phi[\vect{x}(t)]$
and probability measure $P[\vect{x}(t)]$ for the paths.
The relevant paths for Casimir--Polder energies are closed, Gaussian paths
(Brownian bridges) starting and ending at $\vect{x}(0)=\vect{x}(\cT)=\vect{x}_0$;
in discrete form with $N$ discrete path samples, the probability density for these paths in $D-1$ dimensions is 
\begin{align}
  P[\vect{x}(t)]=\,&
  (2\pi\Delta \cT)^{-N(D-1)/2}\,
  e^{-(\vect{x}_{0}-\vect{x}_{N-1})^2/2\Delta \cT}
  \nonumber\\
  &\times
  e^{-(\vect{x}_0-\vect{x}_1)^2/2\Delta \cT}
  \prod_{j=1}^{N-2}e^{-(\vect{x}_{j+1}-\vect{x}_{j})^2/2\Delta \cT},
  \label{eq:pathmeasure}
\end{align}
where $\Delta \cT = \cT/N$. 
To proceed, we will assume the path functional
$\Phi$ to be approximately independent of $\vect{x}_0$
($\partial\Phi/\partial\vect{x}_0\approx 0$).
This is appropriate, for example, in Casimir--Polder calculations where the
dielectric permittivity is constant in the vicinity of the atom
(e.g., as in the case of an atom in vacuum, at a finite distance from a dielectric interface).
In this case, derivatives of the path integral (\ref{eq:general_functional})
with respect to $\vect{x}_0$ act only on the probability density $P$.
For the Gaussian density (\ref{eq:pathmeasure}),
the derivatives along a particular spatial direction
have the form
\begin{equation}
  \frac{\partial^n}{\partial x_{0}^n} P[{x}(t)] = (\Delta \cT)^{-n/2}
  H_n\bigg(\frac{\bar{x}_1-x_0}{\sqrt{\Delta \cT}}\bigg)
  P[{x}(t)],
  \label{eq:Prob-Hermite}
\end{equation}
where we are now indicating the spatial 
dependence in only the relevant
direction for notational simplicity,
$\bar{x}_1 = (x_1+x_{N-1})/2$, and the Hermite polynomials $H_n$ are defined according to the convention
\begin{equation}
  \frac{d^n}{dx^n} e^{-x^2} = (-1)^nH_n(x)\,e^{-x^2}.
\end{equation}
Note that we have written out the derivatives only for one-dimensional paths
to simplify the expressions, but the results here generalize straightforwardly
to multiple spatial dimensions (including multiple derivatives
in one direction as well as cross-derivatives),
in addition to non-Gaussian path measures (e.g., 
for paths corresponding to more general L\'evy processes).

Combining 
Eq.~(\ref{eq:Prob-Hermite}) with the path integral (\ref{eq:general_functional})
gives the expression
\begin{align}
  \frac{\partial^n}{\partial x_{0}^n}
  I &= 
  (\Delta \cT)^{-n/2}
  \biggdlangle 
  H_n\bigg(\frac{\bar{x}_1-x_0}{\sqrt{\Delta \cT}}\bigg)
  \Phi[{x}(t)]\biggdrangle_{{x}(t)} 
  \label{eq:general_functionalderiv}
\end{align}
for the path integral derivatives.
However, this expression is problematic
as a numerical derivative estimate because of the
factor $\Delta\cT^{-n/2}$, which diverges in the continuum limit
$\Delta \cT\longrightarrow 0$. 
This factor indicates large sample variances and cancellations between
paths in this limit, resulting in inefficient numerical computations.

\begin{figure}
\includegraphics{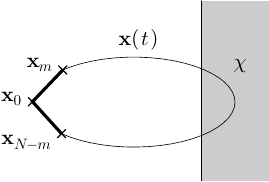}
  \caption{Schematic path showing partial averaging up to the $m$th discrete
    path increment up to and after 
    the source point $\vect{x}_0$.  
    The diagram shows the example geometry of the Casimir--Polder potential
    for an atom in vacuum near a planar interface with a dielectric material
    of susceptibility $\chi$.
  }
  \label{fig:int-by-parts}
\end{figure}

One approach to ameliorating this problematic behavior comes from the observation
that if the path functional $\Phi$ is approximately independent of $\vect{x}_0$,
then in the limit of large $N$, $\Phi$ will also be approximately independent of
neighboring points, such as $\vect{x}_1$, $\vect{x}_{N-1}$, and other nearby points.
In this case, it is possible to carry out the integrals with respect to these
points, to obtain a ``partially averaged'' path integral.
(Note that this method is distinct from other partial-averaging 
methods~\cite{Matacz2000,Predescu03}.)
Carrying out the integrals over $x_1,\ldots,x_{m-1}$
and $x_{N-k+1},\ldots,x_{N-1}$, Eq.~(\ref{eq:general_functionalderiv}) becomes
\begin{align}
  \frac{\partial^n}{\partial x_{0}^n}
  I &\approx
  (2\nu_{mk})^{-n/2}
  \biggdlangle 
  H_n\bigg(\frac{\bar{x}_{mk}-x_0}{\sqrt{2\nu_{mk}}}\bigg)
  \Phi[\vect{x}(t)]\biggdrangle_{{x}(t)},
  \label{eq:general_functionalderiv_partavg}
\end{align}
where $\nu_{mk}$ and $\bar{x}_{mk}$ are defined by
\begin{align}
  \nu_{mk} &:= \frac{mk\Delta \cT}{m+k}\\
  \bar{x}_{mk} &:= \frac{kx_{m} + mx_{N-k}}{m+k},
\end{align}
for $k,m \geq 1$.
The probability density for the paths in the partially averaged expression (\ref{eq:general_functionalderiv_partavg})
is
\begin{align}
  \bar{P}[{x}(t)]&:=\int dx_1\ldots dx_{m-1}dx_{N-k+1}\ldots dx_{N-1}\,P[x(t)]\nonumber\\
  &=e^{-({x}_{0}-{x}_{N-k})^2/2k\Delta \cT}e^{-({x}_{0}-{x}_{m})^2/2m\Delta \cT}
  \nonumber\\
  &\hspace{0.5cm}\times\prod_{j=m}^{N-k-1}e^{-({x}_{j+1}-{x}_{j})^2/2\Delta \cT},
\end{align}
which has ``large steps'' from $x_0$ to $x_m$ and $x_{N-k}$ to $x_0$.
For closed paths, it is generally most sensible to choose $m=k$ to apply
the same partial averaging to the beginning and end of the path, such that
$\nu_m:=\nu_{mm}$ and
$\bar{x}_m:=\bar{x}_{mm}$ simplify to
\begin{align}
  \nu_{m} &= \frac{m\Delta \cT}{2}=\frac{m\cT}{2N} \\
  \bar{x}_{m} &= \frac{x_{m} + x_{N-m}}{2},
  \label{eq:sym-partial-avg}
\end{align}
and the steps at the beginning and end of the path are equally large on average,
as shown schematically (omitting the Brownian nature of the path for visual clarity)
in Fig.~\ref{fig:int-by-parts}.

Since the prefactor in the derivative (\ref{eq:general_functionalderiv_partavg})
now scales as $\smash{\nu_{m}^{-n/2}}$, there is clearly some reduction in the sample
variance in a numeric computation as $m$ increases.
However, the critical point is this: when chosen appropriately, 
this prefactor becomes \textit{independent} of $N$, curing the divergent
sample variance in the continuum limit.
To choose the number $m$ of points to partially average,
the main consideration is that the averaging should not introduce significant error.
Returning to the assumption that $\Phi$ is approximately independent of $\mathbf{x}_0$,
at this point we will have to be somewhat more precise and assume that
$\Phi$ depends on the path $\mathbf{x}(t)$ via a scalar function $V(\mathbf{x})$ in the form
$\Phi[V(\mathbf{x})]$.
Then the key assumption is that $V(\mathbf{x})$ is constant (or approximately constant)
within a radius $d$ of the source point $\vect{x}_0$.
The error in averaging over the point $\vect{x}_m$, while ignoring the
corresponding dependence of $\Phi$, is small so long as there is a small probability 
for $\vect{x}_m$ to leave the constant region.
Assuming $m$ will be chosen to be small compared to $N$, the path from $x_0$ to
$x_m$ (and the corresponding path segment at the end) will behave approximately
as an ordinary Wiener path.
Since
the probability for a one-dimensional Wiener path
to cover a distance $d>0$ from $x_0$ in time $\tau$  is given by 
\begin{equation}
  P_{\text{cross}} = \mathrm{erfc}\left(\frac{d}{\sqrt{2\tau}}\right).
  \label{eq:partialavgerrpre}
\end{equation}
For $\tau\ll d^2$, this expression is bounded above by $e^{-d^2/2\tau}$.
Then to keep the error acceptably small, the crossing probability is chosen
to be below some small tolerance $\varepsilon>0$.  
This condition can be solved to find the total average step time $\tau=m\Delta \cT$ such 
that $P_{\text{touch}}\le\varepsilon$.  
The fraction of the path one can acceptably average over is then given by 
\begin{equation}
  \frac{m}{N} = \frac{d^2}{2\cT\ln (1/\varepsilon)}.
  \label{eq:partialavgerr}
\end{equation}
(In practice we choose $\varepsilon$ to be much smaller than suggested by
this formula.)
Note that $m/N$ is constant even as the path resolution $N$ increases, 
so that the numerical fluctuations,
governed by $\nu_m=m \cT/2N$, are independent of $N$, as required.
A bound for the numerical error then follows from multiplying the tolerance
$\varepsilon$ by the amount over which $V(x)$ varies from $\vect{x}_0$
to the exterior of the constant region.

These apply readily to the computation of derivatives of the Casimir--Polder potential.
For example, the path integral for the Casimir--Polder force becomes
\begin{align}
  \vect{F} =&\, \frac{\hbar c\alpha_0}{4(2\pi)^{D/2}}\int \frac{d\cT}{\cT^{1+D/2}}
  \biggdlangle \frac{2\vect{x}_0-\vect{x}_{m}-\vect{x}_{N-m}}{m\Delta \cT}\nonumber\\
  &\hspace{2.5cm}\times\Big(\langle\epsr\rangle^{-3/2}-[\epsr(\vect{x}_0)]^{-3/2}\Big)
  \biggdrangle_{\vect{x}(t)}
  \label{eq:CPforce}
\end{align}
after partial averaging,
where an appropriate estimator for the path average of the permittivity is
\begin{align}
  \langle \epsr \rangle =&\, \frac{1}{N}\Bigg[m\epsr(\vect{x}_0)+\frac{m}{2}\epsr(\vect{x}_{m})
  +\frac{m}{2}\epsr(\vect{x}_{N-m})
  \nonumber\\
  &\hspace{2.5cm}+\sum_{j=m+1}^{N-m-1}\epsr(\vect{x}_j)\Bigg].
\end{align}
Note that the details of how the points $\vect{x}_{m}$ and $\vect{x}_{N-m}$
enter this estimator are not critical, since for small tolerances $\varepsilon$
the dielectric permittivity at these points should be equal to the source-point permittivity
for the vast majority of paths.
Fig.~\ref{fig:int-by-parts} shows an example of the physical setup for this path integral
for an atom near a planar dielectric surface.
In evaluating this path integral numerically here, the distance scale $d$ is simply the distance
between the atom and the dielectric interface, and a small tolerance $\varepsilon$
guarantees that the points $x_{m}$ and $x_{N-m}$ are unlikely to cross over the interface.

In this differentiated path integral for the force, in principle no 
renormalization term is required---any renormalization term is independent
of the position of the atom, and thus vanishes under the derivative.
That is, the $\epsr(\vect{x}_0)$ term in Eq.~(\ref{eq:CPforce}) does not ultimately
contribute, because it vanishes under the ensemble average for any value of $\cT$.
However, on a path-wise basis this renormalization term is critical in a numerical
implementation of this path integral, because it cuts off
the divergence at $\cT=0$ and dramatically reduces the sample variance of finite
ensemble averages, particularly for small $\cT$.

\subsection{Partial Averages Close to One Body}
\label{section:partial-averaging-close}

The partial-averaging approach is also useful in 
computing derivatives in a geometry with two or more material
bodies, when the source point of the path integral 
is close to one of the bodies.  
This situation applies, for example, 
when computing the two-body stress-energy tensor in the vicinity
of one of the bodies~\cite{Scardiccio2006,Schafer2012,Schafer2016}, 
or when extracting the nonadditive, multibody contribution to the 
Casimir--Polder potential.

\begin{figure}
\includegraphics{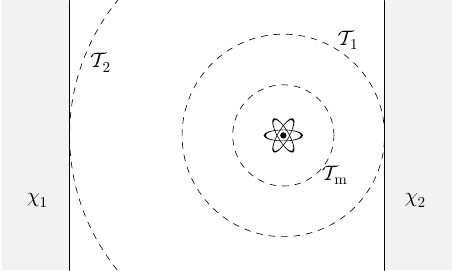}
  \caption{Schematic of an atom between two dielectric half-spaces
    with respective susceptibilities $\chi_1$ and $\chi_2$, illustrating
    the prototype problem for the partial-averaging technique of
    Section~\ref{section:partial-averaging-close} close to one body.  
    The dashed spheres show the typical extent of the 
    paths at various running times $\cT$ (i.e., times over
    which the paths can diffuse).  
    At $\cT_1$ the paths typically touch only the nearer body,
    while at $\cT_2$ they typically touch \textit{both} bodies. 
    The time interval for partial averaging the paths is $\cT_m$.
    The method of Section~\ref{sec:partial_average} requires
    the background dielectric to be approximately uniform in 
    the vicinity of the atom, and thus that $\cT_m\ll \cT_1$ 
    so that the paths are unlikely to touch either interface
    during the partial-averaging time.
    However, because the path integral can be evaluated in the 
    presence of a single, planar interface, the partial averaging
    can be extended such that $\cT_m\sim\cT_1$,
    provided $\cT_m\ll \cT_2$, so that most paths only hit the nearer
    interface during the partial-averaging time.
  }
  \label{fig:atom_2wall}
\end{figure}

As a concrete example, consider differentiating 
the Casimir--Polder potential of 
an atom between two dielectric half-spaces, 
where the atom is close to one interface (Fig.~\ref{fig:atom_2wall}).  
In this problem there are two relevant scales for the path running
time $\cT$: the first time $\cT_1$ that the 
path touches \emph{either} body,
and the first time $\cT_2$ that the path touches \emph{both} bodies.  
For the atom at position $x_0$ between interfaces at 
positions $d_1$ and $d_2$, these
times are on the order of  
$\cT_1\sim (d_1-x_0)^2$ and $\cT_2\sim(d_2-d_1)^2$, respectively.   
If the atom is particularly close to one body,
the dominant contribution to the worldline path integral comes 
from paths where $\cT\sim \cT_2\gg \cT_1$.
The approach of the previous section is limited in this case,
because it relies on the dielectric permittivity being approximately
constant in a neighborhood of the source point, with
the effectiveness of the method increasing with the radius
of this neighborhood.
The method thus only effects partial averaging to a limited
extent, for $\cT_m:= m\Delta \cT\ll \cT_1$.

The partial-averaging method extends to this case 
if there are approximate analytical solutions 
available for the path integral.  
As in the case of the TE-polarization 
path integral~\cite{Mackrory2016},
it is possible to find analytical solutions for path integrals 
corresponding to certain material geometries,
and then to recast more general worldline path integrals in a form
that leverages those solutions.  In particular,
the (un-renormalized) Casimir--Polder energy can be written
as an ensemble average over paths divided into $N$ segments. The
ensemble average in the path integral must then include
an average over the positions of $N$ points that define a discrete
path, as well as averages over all possible path segments connecting
each pair of neighboring points:
\begin{align}
  &V\subCP\supTE(\x0)  =
  \frac{\hbar c\alpha_0}{(2\pi)^{D/2}\sqrt{\pi}\epsilon_0}
  \intzinf \,\frac{d\cT}{\cT^{1+D/2}}\, 
  \intzinf ds\,s^2\,e^{-s^2}\,
  \nonumber\\&\hspace{1mm}\times
  \Biggdlangle
  \prod_{j=0}^{N-1}
  \Bigglinklangle
  \exp\!\Bigg[-\!\frac{s^2}{\cT}\,\int_{\cT_j}^{\cT_{j+1}}\!\!d\tau\,\chi[\mathbf{x}(\tau)]\Bigg]
  \Bigglinkrangle_{\!\!\Delta x_j}
  \Biggdrangle_{\mathbf{x}(\tau)}\!\!\!\!.
  \label{TEpathintexpform2s2repforsubpathavgavg}
\end{align}
Here, the connected-double-angle brackets $\smash{\linklangle\;\linkrangle_{\Delta x_j}}$ denote
an ensemble average over all bridges between $\smash{\mathbf{x}_j}$ and $\smash{\mathbf{x}_{j+1}}$,
 $\cT_j = j\Delta\cT=(j/N)\cT$, 
and the dielectric susceptibility for the two bodies is 
$\chi[\mathbf{x}(\tau)]=\chi_1[\mathbf{x}(\tau)]+\chi_2[\mathbf{x}(\tau)]$.  
With this ordering of brackets, the averages over all path segments 
between points $\smash{\mathbf{x}_j}$ and $\smash{\mathbf{x}_{j+1}}$
occur for a fixed set of discrete points, and then 
the average must be taken over all possible discrete paths
$\vect{x}_j$.
The key point of this section is that  
for simple geometries, such as a single planar interface, 
it is possible to carry out the average over the path segments
analytically. 
In the example of a single, planar interface, the solution
corresponds to the generating function of the sojourn time
of a Brownian bridge
(see Ref~\cite{Mackrory2016}, App.~B).

For sufficiently small time steps ($\Delta \cT\ll \cT_2$), 
the probability for the path to interact with 
both bodies on a given discrete step $j$ is small, so the path integral over bridges is well-approximated by accounting only for the nearest body. 
For path increments $\Delta\mathbf{x}_j$ close to the $\chi_2$ body, 
the path-segment average is then
\begin{equation}
  \Biglinklangle
  e^{-(s^2/\cT)\int d\tau\, (\chi_1+\chi_2)} \Biglinkrangle_{\!\!\Delta \mathbf{x}_j}\approx 
  \Biglinklangle
  e^{-(s^2/\cT)\int d\tau\, \chi_2} \Biglinkrangle_{\!\!\Delta\mathbf{x}_j}.
\end{equation}
The crucial point is that this path integral 
(including the Gaussian probability measure for the path increment),
which we can write as
\begin{align}
  U(\vect{x}_j,\vect{x}_{j+1};t)
  &=\frac{e^{-(\vect{x}_j-\vect{x}_{j+1})^2/2t}}{(2\pi t)^{(D-1)/2}}\Biglinklangle
  e^{-s^2\!\int_0^{t} \!d\tau\, \chi[\vect{x}(\tau)]/\tau} \Biglinkrangle_{\Delta\vect{ x}_j},
  \label{Udef}
\end{align}
corresponds to a propagator for an associated 
diffusion equation~\cite{Feynman1948,Kac1949,Durrett1996,Karatzas1991},
so the solutions between adjacent steps can be composed into
the solution for a larger, combined step.  
The partial averaging can then be carried out 
as in the method of the previous section by composing $m$ steps together.  
To ensure that the error associated with the partial averaging is
small, the probability to touch
the more-distant body in time $\cT_m= m\cT/N$ should
remain below some small tolerance $\varepsilon$.  
In the example of two planar interfaces, this condition has the
explicit form
\begin{equation}
  \frac{m}{N} = \frac{(d_2-x_0)^2}{2\cT\ln (1/\varepsilon)},
  \label{eq:averaging-fraction}
\end{equation}
using the same argument that led to Eq.~(\ref{eq:partialavgerr}).  
In more general material geometries, the analytical path integral 
must also approximate the shape of the nearest body.
For example, a smooth but nonplanar interface (e.g., the surface of
a spherical body) can be approximated
locally by a plane for the purposes of the path-segment average
in a given path increment.
However, in this case, $\cT_m$ is also constrained because
the geometric approximation must hold over the typical extent
of the path up to time $\cT_m$.
For example, in a local planar approximation, 
the distance that paths typically explore in time $\cT_m$
must be small on the curvature scale of the interface.

After averaging, the derivatives are readily computed, 
although in general they will not yield the same Hermite-polynomial
weighting factors as in Eq.~(\ref{eq:general_functionalderiv}):
\begin{align}
  &\hspace{-2mm} \frac{\partial^n}{\partial x_0^n}V\subCP\supTE(\x0) \!  = \!
  \frac{\hbar c\alpha_0}{(2\pi)^{D/2}\sqrt{\pi}\epsilon_0}
  \intzinf \frac{d\cT}{\cT^{1+D/2}} \intzinf ds\,s^2\,e^{-s^2}\,
        \nonumber\\
  &\hspace{5mm}\times
  \int \prod_{k=m}^{N-m}d\vect{x}_k\prod_{j=m}^{N-m-1} U(\vect{x}_j,\vect{x}_{j+1};\Delta \cT)\nonumber\\
  &\hspace{5mm}\times\frac{\partial^n}{\partial x_0^n} [U(\vect{x}_0,\vect{x}_{m};m\Delta \cT)\,U(\vect{x}_{N-m},\vect{x}_0;m\Delta \cT)].
\end{align}
As discussed at the end of the previous section, this expression 
should be renormalized as necessary.
Note that in the path integral here, the derivatives 
act on both the Gaussian path measure and 
the subaverages over path segments
[via the bracketed factor in Eq.~(\ref{Udef})].  
In addition, the functional form of $U$ may vary depending on the 
local path position $\vect{x}_j$, because
it should reflect only the presence of the interface nearest $\vect{x}_j$.  
Although the approximate solution near the source point $\mathbf{x}_0$
of the path is crucial for the partial averaging,
it is not necessary to use such analytical results for the entire path.  For example, one could also use 
the simpler trapezoidal estimator \cite{Mackrory2016} 
for the average of the dielectric function around the remainder
of the path.

The averaging approach described here helps to improve the
sample variance in a numerical average
when the source point of the paths is close to an interface,
even though the functional integral cannot be treated as 
approximately constant, as assumed in the previous method of 
Section~\ref{sec:partial_average}. 
Both this and the previous methods of path-averaging described 
can be thought of as choosing non-uniform time steps 
in the discrete path. 
It is important to note, however, that in most path integrals,
all of the time steps should be vanishingly small for good accuracy;
but in the partially averaged path integrals here, the ``large''
first and last steps should be of fixed time step, even in the limit
as the remaining step sizes become vanishingly small.
Since the derivatives act on the first and last 
path steps, the sample
variance decreases as
their respective time steps increase.  In order to 
improve the statistical convergence of the path average,
the first and last steps should therefore be as large as possible, 
while controlling the errors associated with the
simplified, approximate integrands.
The methods here still allow a high resolution 
for the middle section of the path, in order to 
accurately capture the geometry dependence of the path integral.

\section{Differentiation of Worldline Casimir Energies}
\label{sec:forces}
Path integrals of the
form of Eqs.~(\ref{eq:casimir_energy}) or (\ref{eq:casimir_energy_exp})
correspond to Casimir energies
between macroscopic bodies.
The computation of their derivatives
requires a different
set of methods because they involve an integral over all 
possible source points for the paths.  
To compute forces, curvatures of potentials, and torques 
within this framework, it is necessary to differentiate 
with respect to generalized coordinates, such as distances 
and angles between bodies, in contrast to
the derivatives with respect to path source points that
we considered above.
Here we will develop a general method to handle such 
derivatives of path integrals.

\begin{figure}
\includegraphics[width=\columnwidth]{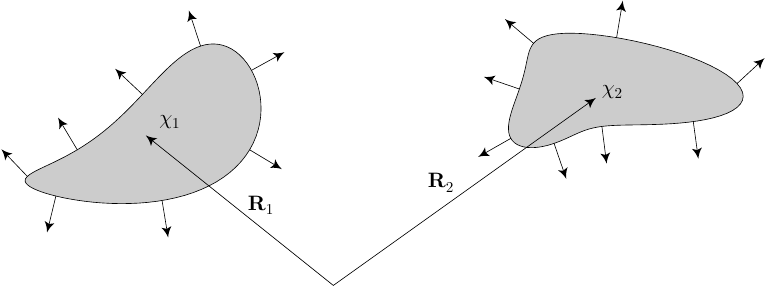}
  \caption{Geometry for interacting dielectric bodies of susceptibility $\chi_j$, centered at
    $\vect{R}_j$ relative to the origin.  
    The normal vectors $\hat{n}_j$ to the surface of the $j$th  body
    are also shown.}
  \label{fig:spud_sketch}
\end{figure}

For the purposes of differentiation,
the critical component of the Casimir-energy path integral has the 
form
\begin{equation}
  E = \biggdlangle \Phi\Big[
   f\Big( \big\langle 
  \chi(\mathbf{x}(t), \mathbf{R}_1,\mathbf{R}_2)  \big\rangle \Big)
  \Big]  \biggdrangle_{\vect{x}(t)},
  \label{eq:spatial_path_integral}
\end{equation}
where $\Phi$ is a linear functional of a function $f$ of
the path average (\ref{eq:path_average}) of the 
susceptibility $\chi(\mathbf{r})$,
and where $\mathbf{R}_1$ labels a generalized coordinate 
of the first body (or collection of bodies),
and $\mathbf{R}_2$ labels a generalized coordinate 
of the second body (or collection thereof).
For example, a typical arrangement is depicted in
Fig.~\ref{fig:spud_sketch}, where
uniform dielectric bodies are
separated by vacuum, although the treatment here
may be straightforwardly generalized to nonuniform dielectric media.
In this case, the susceptibility $\chi(\vect{r})$ is given
generally by 
\begin{equation}
  \chi(\vect{r}) = \sum_j\chi_j\Theta[\sigma_j(\vect{r}-\vect{R}_j)],
\end{equation}
where $\chi_j$ is the dielectric susceptibility of body $j$;
$\sigma_j(\vect{r})=0$ 
defines the surface of the $j$th body, with $\sigma_j>0$ and $\sigma_j<0$ 
on the interior and exterior of the body,
respectively; and $\vect{R}_j$ is the center of the $j$th body.  
The integral still averages over the paths $\vect{x}(t)$ via the
path average.
Recall that $\Phi$ depends implicitly on $\cT$ via the path statistics,
because $\vect{x}(0)=\vect{x}(\cT)=\vect{x}_0$.

\subsection{Force}
\label{sec:force}

The force on a body follows from the gradient of the Casimir energy,
where the derivatives are taken with respect to the body's 
position.
For example, the force on body $1$ is
given by differentiating the path integral in Eq.~(\ref{eq:spatial_path_integral})
with respect to the body position $\mathbf{R}_1$:
\begin{equation}
  \mathbf{F}_1 = 
  - \nabla_{\mathbf{R}_1} E
  =- \,\biggdlangle 
  \frac{\partial\Phi}{\partial  f}\,
  \nabla_{\mathbf{R}_1}  f
  \biggdrangle_{\vect{x}(t)}.
\end{equation}
In this expression,
$\partial\Phi/\partial f$ is independent of $f$
(since $\Phi$ is a linear function of $f$).
In the case where $f$ is exponential, 
we can divide the path average into two steps:
first, averaging over the $N$ discrete, uniform steps 
$\Delta\mathbf{x}_j := \mathbf{x}_{j+1}-\mathbf{x}_j$,
and second, averaging over all the subpaths between $\mathbf{x}_j$
and $\mathbf{x}_{j+1}$:
\begin{equation}
  \mathbf{F}_1 
  =- \,\Biggdlangle 
  \frac{\partial\Phi}{\partial f}\,
  \nabla_{\mathbf{R}_1}\prod_{j=0}^{N-1}
   \biglinklangle f\biglinkrangle_{\Delta\mathbf{x}_j}
  \Biggdrangle_{\vect{x}_n}.
\end{equation}
Here, the connected-double-angle brackets
$\smash{\linklangle~\linkrangle_{\Delta\mathbf{x}_j}}$
denotes an average over all such bridges between $\mathbf{x}_j$
and $\mathbf{x}_{j+1}$, as in Sec.~\ref{section:partial-averaging-close},
and the subscript $\mathbf{x}_n$ indicates an average over all discrete paths
of $N$ steps.
Then applying the derivative to the product of subpaths,
\begin{equation}
  \mathbf{F}_1 
  =- \,\Biggdlangle 
  \frac{\partial\Phi}{\partial f}\,
   f \sum_{j=0}^{N-1}
   \frac{\nabla_{\mathbf{R}_1}\biglinklangle f\biglinkrangle_{\Delta\mathbf{x}_j}}{\biglinklangle f\biglinkrangle_{\Delta\mathbf{x}_j}}
  \Biggdrangle_{\vect{x}_n}.
  \label{expform1deriv}
\end{equation}
There are typically $N$ terms here, but if 
$\smash{\linklangle f\linkrangle_{\Delta\mathbf{x}_j}}$ is associated with a sharp surface, only $\smash{O(N^{1/2})}$ of them
will contribute to the sum.
Again, this form of the path integral is specific to an exponential $f$;
for example, Eq.~(\ref{eq:casimir_energy_exp}) is of the form
\begin{align}
  &f\big(\langle\chi\rangle\big)
  =e^{-s^2\cT(1+\langle \chi\rangle)/2c^2}\\
  &\nonumber\Phi = -\frac{\hbar}{(2\pi)^{5/2}}\int_0^\infty\!\!ds
  \int_0^\infty \!\frac{d\cT}{\cT^{5/2}}\int d\mathbf{x}_0\,f(\langle\chi\rangle),
\end{align}
and works with this method.
Other path integrals, for example of the form of Eq.~(\ref{eq:casimir_energy}),
\begin{align}
  &f(\langle\chi\rangle)=(1+\langle\chi\rangle)^{-1/2}
  \\&\nonumber
  \Phi = -\frac{\hbar c}{8\pi^{2}}\int_{0}^{\infty}\!
  \frac{d\cT}{\cT^3}\int d\vect{x}_0 
  \,f(\langle\chi\rangle),
\end{align}
can be handled using the alternate form for the force,
\begin{equation}
  \mathbf{F}_1 
  =- \,\Biggdlangle 
  \frac{\partial\Phi}{\partial f}\,
  f'
  \sum_{j=0}^{N-1}\nabla_{\mathbf{R}_1}
   \biglinklangle\chi\biglinkrangle_{\Delta\mathbf{x}_j}
  \Biggdrangle_{\vect{x}_n},
\end{equation}
where $\cT_j=j\Delta\cT=j\cT/N$.
Qualitatively, the Casimir force on a body arises from 
paths that pierce its surface, with a vector
path weight
in the direction of the surface normal at the path source point. 
Although the surface of an arbitrary body involves surface normals
pointing in all directions, each surface normal obtains a 
different geometry-dependent
weight via the path ensemble.  The result is, in general,
a nonzero net force.

As in Sec.~\ref{sec:partial_average}, a
partial-averaging procedure accelerates convergence.
If one surface is located at $d_1$, then the condition is
\begin{equation}
  (x_{j+1}-d_1)(x_j-d_1)\leq\varepsilon
  \label{eq:partialavgerr2plane}
\end{equation}
for a positive tolerance $\varepsilon$ [typically $10^4/(N\sqrt{\cT})$; note that
setting $\varepsilon=0$ simply detects a surface crossing].
When this condition is satisfied for any boundary,
the solution advances forward by $m>1$ points, and the estimate of the
function $\linklangle f\linkrangle_{\Delta\mathbf{x}_j}$
or the susceptibility 
$\linklangle \chi\linkrangle_{\Delta\mathbf{x}_j}$
is calculated with respect to this larger path length.

\subsection{Potential Curvature}

This method can be easily extended to the second derivative of the worldline energy, which 
yields the potential curvature,
\begin{equation}
  C_{ij} := (\hat{r}_i\cdot \nablaR1)(\hat{r}_j\cdot \nablaR1)E,
\end{equation}
in basis $\hat{r}_i$.
Applying this to the path integral (\ref{eq:spatial_path_integral}) gives
\begin{equation}
  C_{ij} = 
  \Biggdlangle 
  \frac{\partial\Phi}{\partial f}\,
  (\hat{r}_i\cdot \nablaR1)
  (\hat{r}_j\cdot \nablaR1)
  \prod_{k=0}^{N-1}
   \biglinklangle f\biglinkrangle_{\Delta\mathbf{x}_k}
  \Biggdrangle_{\vect{x}_n},
\end{equation}
and applying the derivatives in the case where $f$ is exponential yields
\begin{align}
  C_{ij} =& 
  \Biggdlangle 
  \frac{\partial\Phi}{\partial f}\,f
  \Bigg[
  \sum_{\substack{\ell, k=0\\ \ell\neq k}}^{N-1}
   \frac{\hat{r}_i\cdot \nablaR1
  \biglinklangle f\biglinkrangle_{\Delta\mathbf{x}_\ell}}{\biglinklangle f\biglinkrangle_{\Delta\mathbf{x}_\ell}}
  \frac{\hat{r}_j\cdot \nablaR1
   \biglinklangle f\biglinkrangle_{\Delta\mathbf{x}_k}}{\biglinklangle f\biglinkrangle_{\Delta\mathbf{x}_k}}
   \\ &\hspace{10mm}+
   \sum_{k=0}^{N-1}
   \frac{(\hat{r}_i\cdot \nablaR1)(\hat{r}_j\cdot \nablaR1)
   \biglinklangle f\biglinkrangle_{\Delta\mathbf{x}_k}}
   {\biglinklangle f\biglinkrangle_{\Delta\mathbf{x}_k}}
  \Bigg]
  \Biggdrangle_{\vect{x}_n}.
  \nonumber
\end{align}
However, it is possible to recast one derivative with 
respect to the other surface,
\begin{equation}
  C_{ij} = 
  -\,\Biggdlangle 
  \frac{\partial\Phi}{\partial f}\,
  (\hat{r}_i\cdot \nablaR2)
  (\hat{r}_j\cdot \nablaR1)
  \prod_{k=0}^{N-1}
   \biglinklangle f\biglinkrangle_{\Delta\mathbf{x}_k}
  \Biggdrangle_{\vect{x}_n},
\end{equation}
in which case,
applying the derivatives, in the case where $f$ is exponential, and
making the approximation $\Delta\mathbf{x}_j\ll d$
for well separated bodies, where $d$ characterizes the body separation,
we obtain
\begin{align}
  C_{ij} \approx&
  -\,\Biggdlangle 
  \frac{\partial\Phi}{\partial f}\,f
   \Bigg[
  \sum_k 
   \frac{\hat{r}_i\cdot \nablaR2\biglinklangle f\biglinkrangle_{\Delta\mathbf{x}_k}}
     {\biglinklangle f\biglinkrangle_{\Delta\mathbf{x}_k}}
    \Bigg]
    \\ &\hspace{15mm}\times
    \Bigg[
  \sum_k 
   \frac{\hat{r}_j\cdot \nablaR1\biglinklangle f\biglinkrangle_{\Delta\mathbf{x}_k}}
     {\biglinklangle f\biglinkrangle_{\Delta\mathbf{x}_k}}
   \Bigg]
  \Biggdrangle_{\vect{x}_n}.
  \nonumber
  \label{expform2deriv}
\end{align}
In the general case, where $f$ is any function,
in the same approximation we obtain
\begin{align}
  C_{ij} \approx&
  -\,\Biggdlangle 
  \frac{\partial\Phi}{\partial f}\,f''
   \sum_{k}\hat{r}_i\cdot\nabla_{\mathbf{R}_2}
   \biglinklangle\chi\biglinkrangle_{\Delta\mathbf{x}_k}
    \\ &\hspace{15mm}\times
   \sum_{\ell}\hat{r}_j\cdot\nabla_{\mathbf{R}_1}
   \biglinklangle\chi\biglinkrangle_{\Delta\mathbf{x}_\ell}
  \Biggdrangle_{\vect{x}_n}.
  \nonumber
\end{align}
Partial averaging is then carried out as in 
Sec.~\ref{sec:force}.

\subsection{Torque}
The Casimir torque on a body follows in a similar way by considering 
the energy shift due to a small rotation.
Specifically, consider an infinitesimal rotation of body 1 about the point $\vect{R}_1$,
with axis $\hat{m}$, and
through angle $\phi$:
\begin{equation}
  \chi(\vect{r}) = \chi_1\Theta\big\{\sigma_1[\mathcal{R}(\phi)(\vect{r}-\vect{R}_1)]\big\}
  +\chi_2\Theta[\sigma_2(\vect{r}-\vect{R}_2)].
\end{equation}
For a small rotation angle, the rotation matrix has the form
\begin{equation}
  \mathcal{R}_{ij}(\phi) = \delta_{ij} - m_k\epsilon_{ijk}\phi +O(\phi^2),
\end{equation}
where $\delta_{ij}$ is the Kronecker delta 
and $\epsilon_{ijk}$ is the Levi--Civita symbol;
repeated indices are summed 
throughout this section.
The torque for a rotation about axis $\hat{m}$ can be written as $K_m:=\hat{m}\cdot\vect{K}=-\partial_\phi E$.
The $\phi$-derivative acts only on the path-averaged dielectric part of the energy integral,\begin{align}
  \partial_\phi\langle\chi\rangle&=
  \chi_1\big\langle \partial_\phi\mathcal{R}_{ij}(\phi)[\vect{x}(t)-\vect{R}_1]_j[\hat{r}_i\cdot\nabla\Theta(\sigma_1)]\big\rangle\nonumber\\
  &=-\chi_1\big\langle m_k\epsilon_{kij}[\vect{x}(t)-\vect{R}_1]_j[\hat{r}_i\cdot\nabla\Theta(\sigma_1)]\big\rangle\nonumber\\
  &=\chi_1\hat{m}\cdot\big\langle [\vect{x}(t)-\vect{R}_1]\wedge\nabla\Theta(\sigma_1)\big\rangle,
\end{align}
where in the last step we introduced the usual vector cross-product 
$(\vect{a}\wedge\vect{b})_i=\epsilon_{ijk}a_jb_k$.  
Substituting this into Eq.~(\ref{eq:spatial_path_integral}), 
the resulting path integral
for the torque, after dropping the arbitrary axis $\hat{m}$, is 
\begin{align}
  \vect{K} &= 
  \,\Biggdlangle 
  \frac{\partial\Phi}{\partial f}\,
  f'\chi_1 
  \sum_{j=0}^{N-1}
   \biglinklangle
   \nabla\Theta(\sigma_1)
     \wedge [\mathbf{x}(t)-\mathbf{R}_1]
   \biglinkrangle_{\Delta\mathbf{x}_j}
  \Biggdrangle_{\vect{x}(t)}
  \label{eq:torque}
\end{align}
This expression for the torque has the intuitive interpretation where
the contribution from each surface element is 
the cross-product of the displacement
vector from the rotation center to the surface
point with the Casimir force density associated with that surface element.

\section{Numerical Results}
\label{sec:numerical_results}

In this section we demonstrate the feasibility and good performance of the above methods
for numerically differentiating worldline path integrals.
As in previous work~\cite{Mackrory2016},
the basic Casimir--Polder calculation of an atom interacting with a planar dielectric
interface, along with the Casimir energy of two thin,
parallel, planar dielectric membranes,
will serve as the test cases for these methods.

\subsection{Partial Averaging}
\label{sec:partial_averaging_numerical}

\begin{figure}[tb]
  \begin{center}
\includegraphics[width=\columnwidth]{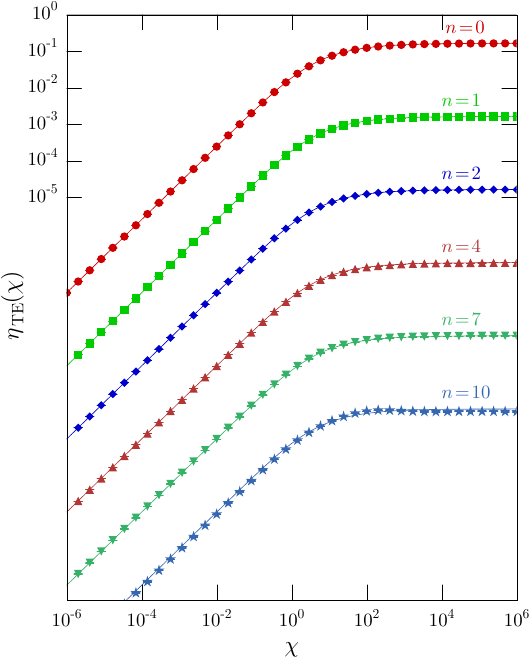}
  \end{center}
  \caption
  {
    Numerical evaluation of the path integral (\ref{eq:CP_energy})
    for the (normalized, dimensionless) Casimir--Polder potential
    of an atom near a dielectric half-space,
    as a function of the (dimensionless) dielectric susceptibility $\chi$
    (data shown as circles).
    The remaining curves, each successively offset by a factor of $10^{-2}$
    for clarity, show spatial derivatives of the potential
    according to the method of Eq.~(\ref{eq:general_functionalderiv_partavg}),
    for $n=1$ (squares), $2$ (diamonds), $4$ (triangles), $7$ (inverted triangles),
    and $10$ (stars) derivatives.  
    The solid lines give the analytic result (\ref{etaTEanalytic})
    for comparison in each case.  Error bars delimit one standard error.
    Details of the calculations are given in the text, but the computational
    difficulty increases with increasing $n$, and accuracy noticeably suffers
    by the $n=10$ case.
    The same random numbers are used for all the points in the same curve.
    \label{fig:upto10derivs}
  }
\end{figure}

Fig.~\ref{fig:upto10derivs} demonstrates the utility
of the Hermite-weighting 
method of Eq.~(\ref{eq:general_functionalderiv_partavg}) for computing
derivatives of path integrals.
The plot shows the results of the numerical evaluation
of the path integral (\ref{eq:CP_energy}) for the Casimir--Polder potential
of an atom near a dielectric half-space, 
along with the computed spatial derivatives
up to
the tenth.  Since the potential and its derivatives
are normalized to the potential in the perfect-conductor limit,
the analytic result is distance-independent, and has the same form in each case, given
by the expression
\begin{equation}
  \eta\subTE(\chi)=
  \frac{1}{6} + \frac{1}{\chi} -\frac{\sqrt{1+\chi}}{2\chi}-\frac{\sinh^{-1}\!\sqrt{\chi}}{2\chi^{3/2}},
  \label{etaTEanalytic}
\end{equation}
as discussed in Ref.~\cite{Mackrory2016} (while not completely
realistic \cite{Babb2004}, this model suffices for a numerical test).

Some aspects of the calculations in this figure are worth discussing here.
These computations nominally employed
$N=10^5$ points per path.  However, at finite $N$, the
error analysis in Ref.~\cite{Mackrory2016} indicates that numerical accuracy suffers,
particularly at large values of the susceptibility $\chi$,
because the finitely sampled path does not have the same ``extent'' as
the same path in the continuum limit. To help maintain accuracy
at large $\chi$, the paths employ some limited, finer sampling:
the two path intervals on either side of the point closest to the dielectric
interface are each sampled with $10^3$ more points
(see App.~\ref{app:openbridge} for the algorithm to generate these sub-paths).
This strategy substantially alleviates the major source of sampling error at large $\chi$, while
negligibly impacting the computational effort.
In terms of the number $m=k$ of partial averages, we make the simplistic choice
of choosing a \textit{fixed} value of $m$ for each calculation
($m=10$, 
20,
200,
200, and
500
partial averages were employed
for $n=1$, 2, 4, 7, and 10 derivatives, respectively),
despite the indication of Eq.~(\ref{eq:partialavgerr}) that the number of
partial averages
should be chosen to be inversely proportional to $\cT$ to maintain
a controlled error.
This is justified by noting that the partial averages can be chosen
to have small errors for small values of $\cT$, and the effects of the
resulting larger error
for large values of $\cT$ will be suppressed by the factor $\cT^{-3}$
in the path integral (\ref{eq:CP_energy}).
(For high-accuracy evaluation of the path integral, the partial 
averages should instead be chosen in a $\cT$-dependent way.)

Another major numerical challenge that arises in Fig.~\ref{fig:upto10derivs}
in the case of high-order
derivatives comes from the Hermite-polynomial factor
in Eq.~(\ref{eq:general_functionalderiv_partavg}).
One issue here is that the Hermite polynomial takes on large
values for large values of $|\bar{x}_m-x_0|$.  But the Gaussian
path measure in the path integral only generates such large values rarely.
This means that, with ordinary Gaussian paths, rare events make 
large contributions to the ensemble, leading to poor statistical
accuracy in the numerical computations.
In these calculations, we employed the direct solution of sampling
$\bar{x}_m$ from the Hermite--Gaussian density
\begin{equation}
  P_{H_n}(\bar{x}_m)=
  \frac{\eta_{n}}{\sqrt{\pi m\Delta \cT}}
  \left|H_n\!\left(\frac{\bar{x}_{m}-x_0}{\sqrt{m\Delta \cT}}\right)\right|
  e^{-(\bar{x}_{m}-x_0)^2/m\Delta \cT},
  \label{eq:Hmstep}
\end{equation}
where the factor
\begin{equation}
  \eta_{n}^{-1}
  :=\frac{2}{\sqrt{\pi}}\intzinf dx\,\left|H_n(x)\right|\,
  e^{-x^2}
  \label{eq:Hmstepnorm}
\end{equation}
normalizes the probability distribution.
Then a Gaussian deviate $\delta x_m$ of zero mean and variance
$m(1-2m/N)\Delta \cT/2$ allows the
first and last path steps to be generated according to
$x_m=\bar{x}_{m}+\delta x_{m}$ and $x_{N-m}=\bar{x}_{m}-\delta x_{m}$,
and the remainder of the path is generated as a Brownian
bridge running from $x_m$ to $x_{N-m}$ in time $(N-2m)\Delta \cT$
(see App.~\ref{app:openbridge} for the bridge-generating algorithm).
With these modified paths,
the path average (\ref{eq:general_functionalderiv_partavg}) should be
modified to read
\begin{align}
  \frac{\partial^n}{\partial x_{0}^n}
  I &\approx
  \eta_{n}^{-1}(m\Delta \cT)^{-n/2}
  \,\biggdexpct{\Phi[\vect{x}(t)]\,\sgn\big[H_n(\bar{z})\big]}_{H_n},
  \label{eq:general_functionalderiv_partavg_withHmpaths}
\end{align}
where $\bar{z}:=(\bar{x}_m-x_0)/\sqrt{m\Delta\cT}$.
Because the Hermite polynomial can take either sign,
a great deal of cancellation occurs in 
Eq.~(\ref{eq:general_functionalderiv_partavg_withHmpaths}) among the paths,
which can again lead to poor statistical performance for high-order derivatives,
when the Hermite polynomial has multiple regions of opposing sign.
To combat this, for a given path, the path functional can be averaged
over multiple values of $\bar{x}_m$ (holding the rest of the path fixed) 
to effect this cancellation on a pathwise basis.
The calculations in Fig.~\ref{fig:upto10derivs},
used averages over
1, 
5,
100,
5000, and
5000
values of $\bar{x}_m$ per path
for $n=1$, 2, 4, 7, and 10 derivatives, respectively,
where the $\bar{x}_m$ were sampled 
from a sub-random (1D van der Corput) sequence to enhance the
accuracy of this subaverage.

The numerical performance in Fig.~\ref{fig:upto10derivs}
is generally quite good, although performance suffers for high-order
derivatives.  
All of the data here represent statistical averages over $10^8$ paths
(except for $n=10$ derivatives, which averages over $10^9$ paths).
The points at a given derivative order $n$ were computed
using the same random numbers, so the points within a curve are not statistically
independent. The error bars (delimiting one standard error) 
are invisibly small on the plot, and amount to a relative statistical
error of at most
$0.2\%$,
$0.5\%$,
$0.7\%$,
$0.8\%$,
$2\%$, and
$2\%$
for $n=0$, 1, 2, 4, 7, and 10 derivatives, respectively.
The observed accuracy of the results is comparable to the standard error
in each case except for the 10th derivative, where the observed error
is at worst 14\%, despite the extra computational effort.
However, note that a comparable computation of the 10th derivative 
via the finite-difference method would be completely unfeasible.
For an effective resolution of $N=10^8$ points per path
(with $N=10^5$ and subsampling of $10^3$ points as described above),
a length scale of the order $N^{-1/2}=10^{-4}$ is appropriate for the finite
difference, but this leads to a standard deviation in the path values
of the order of $10^{40}$ relative to the mean.

\subsection{Convergence: Finite Differences vs.\ Partial Averages}

\begin{figure}
\includegraphics[width=\columnwidth]{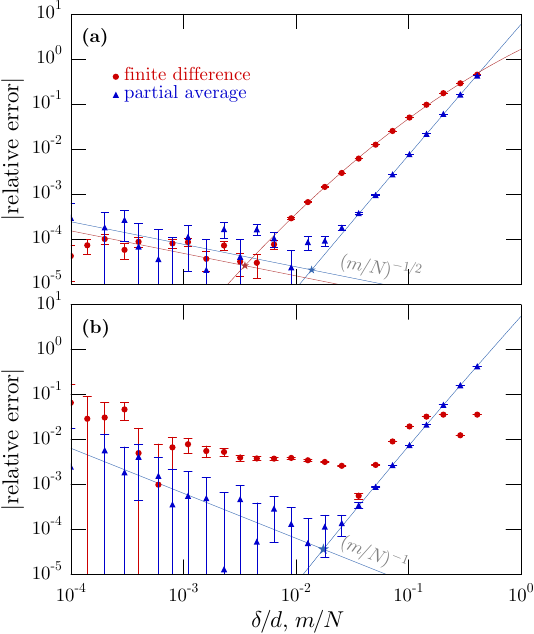}
  \caption{
  Magnitude of the relative error in the computation of (a)
  one spatial derivative and (b) two derivatives 
  of the atom--surface
  Casimir--Polder potential, for a surface dielectric susceptibility
  $\chi=100$.
  The plots compare the performance of the finite-difference
  (circles)
  and partial-averaging (triangles) methods.
  The finite-difference data are plotted with respect to the
  spatial displacement $\delta$, normalized to the
  atom--surface distance $d$, while the
  partial-averaging results are plotted as a function
  of the averaging fraction $m/N$, where $m$ appears
  in Eq.~(\ref{eq:sym-partial-avg}) and $N$ is the number
  of points per discrete path.
  The calculations employed paths with $N=10^4$ points per path,
  averaging over $10^{11}$ paths.
  Error bars delimit one standard error.
  The lines are fits to the data 
  (except for the second derivative in the finite-difference case),
  and the intersection (marked by a star) gives the estimated best 
  performance of the method.
  Data points within each method are statistically independent.
  }
\label{fig:convergence}
\end{figure}

\begin{figure}
\includegraphics[width=\columnwidth]{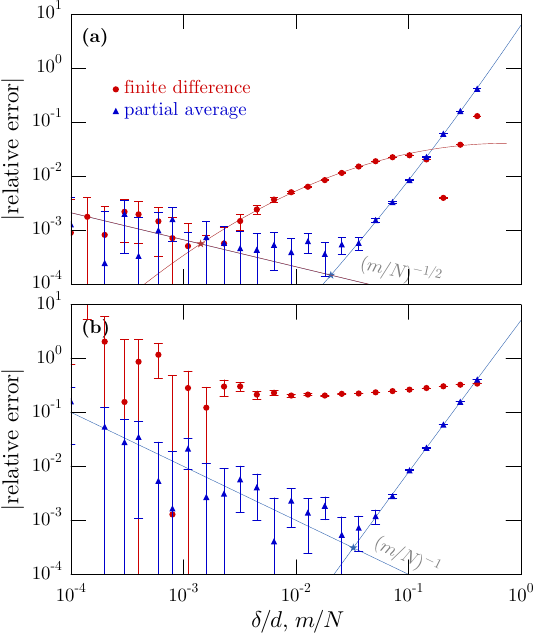}
  \caption{
  Magnitude of the relative error in the computation of (a)
  one spatial derivative and (b) two derivatives 
  of the atom--surface
  Casimir--Polder potential.
  The calculations here are similar to those in
  Fig.~\ref{fig:convergence}, except here the surface
  is a perfect conductor ($\chi\longrightarrow\infty$).
  Also, the calculations here employed paths 
  with $N=10^6$ points per path (with subpaths of $10^5$ points
  on either side of the point closest to the surface),
  averaging over $10^9$ paths.
  Other details are as in Fig.~\ref{fig:convergence}.
  }
\label{fig:convergence-chi-inf}
\end{figure}

To study the numerical error of the partial-averaging 
method in more detail,
we performed the same calculations as in the previous
section for one and two spatial
derivatives of the 
Casimir--Polder potential for an atom near a planar interface, 
while varying the number $m$ of partial averages
(Figs.~\ref{fig:convergence} and \ref{fig:convergence-chi-inf}).
To simplify the analysis, $m$ is held constant (independent of
$\cT$), ignoring the error-control criterion (\ref{eq:partialavgerr}).
For comparison, the plots also show the corresponding errors
when computing the derivatives via finite differences.
The first derivative employs the second-order, 
centered-difference formula 
$V'\subCP(d)=[V\subCP(d+\delta)-V\subCP(d-\delta)]/2\delta
+O(\delta^2)$,
where $\delta$ is a small spatial displacement.
The second derivative exploits the analogous difference formula
$V''\subCP(d)=[V\subCP(d+\delta)-2V\subCP(d)+V\subCP(d-\delta)]/\delta^2
+O(\delta^2)$.
In implementing the finite-difference methods,
it is important to apply these difference formulae on a path-wise
basis (or equivalently, apply the difference formulae on
ensemble averages over equivalent path samples) for this
method to produce any kind of reasonable result.
In applying the finite-difference method, the numerical details 
(path generation,
calculation of path averages, etc.)
otherwise match those of previous work (see 
Ref.~\cite{Mackrory2016}, Section~V).
Although the error for both the partial-averaging and
finite-difference methods behaves similarly in the plots,
note that the quantities $m/N$ and $\delta/d$ from the two methods
are not directly comparable. In the partial-averaging
method, $m$ sets a length scale $\sqrt{m\Delta\cT}=\sqrt{(m/N)\cT}$
that varies with $\cT$ in the integral in Eq.~(\ref{eq:CP_energy}).
However, it is customary practice for the finite-difference method
to use a displacement $\delta$ independent of the details
of the rest of the calculation, so $\delta$ represents a fixed
length scale, independent of $\cT$.  Nevertheless, the plots
provide a useful visual comparison of the performance of the two methods.

The performance of the two methods
for a dielectric susceptibility $\chi=100$
(Fig.~\ref{fig:convergence}) is similar in minimizing the
relative error in the case of one derivative.
However, the partial-averaging method performs
clearly better in the two-derivative case.
Although $\chi\gg 1$ in this computation, for numerical purposes
this calculation is in a ``small-$\chi$'' regime because
$N\gg\chi$~\cite{Mackrory2016}.
In terms of the path integral, this means that, in the
first-derivative case, when
comparing two slightly displaced copies of the same path
that overlaps the dielectric interface, the difference in the path-average
functional $\langle\epsr\rangle^{-3/2}$ from the path integral
(\ref{eq:CP_energy}) will be small, of the order $1/N$ for
sufficiently small displacements.
Finite differences tend to work well when operating on
smooth functions, and this is exactly what
Fig.~\ref{fig:convergence}(a) shows: despite the noisy nature of the
paths, the path integral in this regime is relatively well
behaved, allowing the finite differences to exhibit
performance comparable to that of the partial-averaging method.
The clear separation between the methods in 
Fig.~\ref{fig:convergence}(b) highlights the better 
performance of the partial-averaging method
for derivatives beyond the first.

Figure~\ref{fig:convergence} also shows fits to the data
(the range of each fit is shown in App.~\ref{app:endpoints}, Table~\ref{table:table1}).
On the side of small $(\delta/d)$ or $(m/N)$, the data are fit
to $a(\delta/d)^{-1/2}$ or $a(m/N)^{-1/2}$ in 
Fig.~\ref{fig:convergence}a,
and to $a(m/N)^{-1}$ in
Fig.~\ref{fig:convergence}b.
This reflects the $(\delta/d)^{-n/2}$ or $(m/N)^{-n/2}$ scaling
expected of the statistical error in the case of $n$ derivatives
[particularly in the latter case, where 
Eq.~(\ref{eq:general_functionalderiv_partavg})
has a prefactor that scales in this way].
On the side of large $(\delta/d)$ or $(m/N)$,
the logarithms of the data (on both axes) are fit to a second-order polynomial.
The crossings of the corresponding curves are marked by stars,
which give the best error expected from the method.
The relative errors in the first-derivative methods are both around 
$2\times 10^{-5}$ and the relative error in the second-derivative partial-average
method is around $4\times 10^{-5}$.
The first-derivative error also 
exhibits a spurious dip (actually a zero), as noted above,
for $\delta/d$ near $0.2$. However, this does not represent a reliable
error minimum in a more general calculation.  Rather, the
asymptotic $(\delta/d)^2$ scaling only just becomes visible in a relatively
narrow interval (near the center of the interval marked between
$\delta/d=10^{-3}$ and $10^{-1}$) before the statistical error takes over.
This region represents the true best-case error of this method
in this calculation.

The difference in performance between the methods is even more clear
in the perfect-conductor ($\chi\longrightarrow\infty$)
limit (Fig.~\ref{fig:convergence-chi-inf}).
In both the first- and second-derivative
cases the partial-averaging method achieves substantially
better accuracy than the finite-difference method.
The contrast is particularly striking
for the second derivative, where the finite difference
achieves a minimum error of around 20\%, making the method
effectively useless in this regime.
(Note that in the finite difference of the first derivative,
there is a spurious zero near
$\delta/d=0.2$, as discussed above.)
The poor performance of the finite difference is consistent with the
argument in the introduction.
In the perfect-conductor regime, 
slightly shifted paths can produce path-functional values
that differ by much more than $\delta$---the path-average
functional $\langle\epsr\rangle^{-3/2}$ from the path integral
(\ref{eq:CP_energy}), after renormalization, leads
to a null value for paths that do not overlap the interface,
but a value of unity for overlapping paths.
This leads to large statistical errors in a finite, numerical
path average, because finite differences lead
to path-functional values switching between zero and $\delta^{-n}$
for $n$ derivatives.  Also, as is evident in the data, the systematic
errors become relatively large in this regime as well, especially
in the case of the second derivative.
In fact, because we know from prior work~\cite{Mackrory2016}
that the calculations in the perfect-conductor limit are subject
to larger systematic errors than small-$\chi$ (i.e., $N\gg\chi$)
calculations, we used substantially more points per path in the large-$\chi$
calculations ($N=10^6$ compared to $N=10^4$ for $\chi=100$,
with the addition here of
sub-path sampling, using $10^5$ points per interval near
the path point closest to the surface, as described in the previous
section; to keep the numerical effort comparable in the two cases,
the large-$\chi$ results average over $10^9$ paths, compared
to the $10^{11}$ paths in
the $\chi=100$ case).

The same curve fits were applied in Fig.~\ref{fig:convergence-chi-inf}
(fit ranges in App.~\ref{app:endpoints}, Table~\ref{table:table1}).
For the first derivative, the finite-difference best relative error
is $5\times 10^{-4}$, while the partial-average best relative error is
$1\times 10^{-4}$, a difference of a factor of 5.
In the second-derivative case, the partial-average method performs
much better than the finite-difference, with a relative error
of $3\times 10^{-4}$.

At this point it is worth reiterating one of the main motivations
for developing the partial-averaging method for numerical differentiation.
The computation of the Casimir--Polder potential in the case of 
transverse-magnetic (TM) polarization involves the second derivative of a path integral
with a path-integral potential that is highly singular at a dielectric
interface.  Indeed, the TM potential at an interface between finite-$\chi$
dielectric materials turns out to be far more 
pathological than the above behavior of the TE path integral
for an interface in the 
$\chi\longrightarrow\infty$ limit.
This pathological behavior is problematic when applying
the finite-difference method to the
numeric differentiation of the TM
path integral.  However, as we will show in a future publication,
the partial-averaging method is a serviceable alternative method
for computing the TM-polarization potential.

\subsection{Casimir Force and Curvature}\label{sec:casforcecurve}

\begin{figure}[tb]
  \begin{center}
\includegraphics[width=\columnwidth]{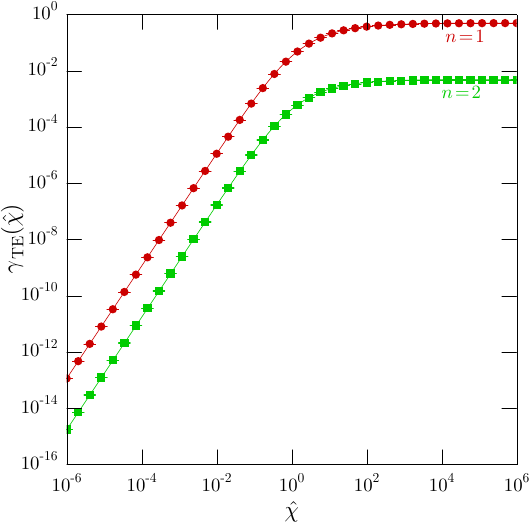}
  \end{center}
  \caption
  {
    Numerical evaluation of the derivatives of the
    path integral (\ref{eq:casimir_energy_exp})
    for the (normalized, dimensionless) Casimir potential of two dielectric
    thin planes,
    as a function of the (dimensionless) susceptibility parameter $\hat{\chi}$
    (data shown as circles).
    The first ($n=1$) derivative is shown (circles) along with the
    second ($n=2$) derivative (squares, offset by a factor $10^{-2}$ for clarity).
    The solid lines give the derivatives of the 
    analytic result (\ref{gammaTE}) for comparison in each case.  
    The calculation used $10^9$ paths of $10^5$ points per path.
    Error bars delimit one standard error (on the order of a few times $10^{-4}$
    to a few times $10^{-3}$ in this plot).
    The same random numbers are used for the points in each curve.
    \label{fig:upto2derivs}
  }
\end{figure}

\begin{figure}[tb]
  \begin{center}
\includegraphics[width=\columnwidth]{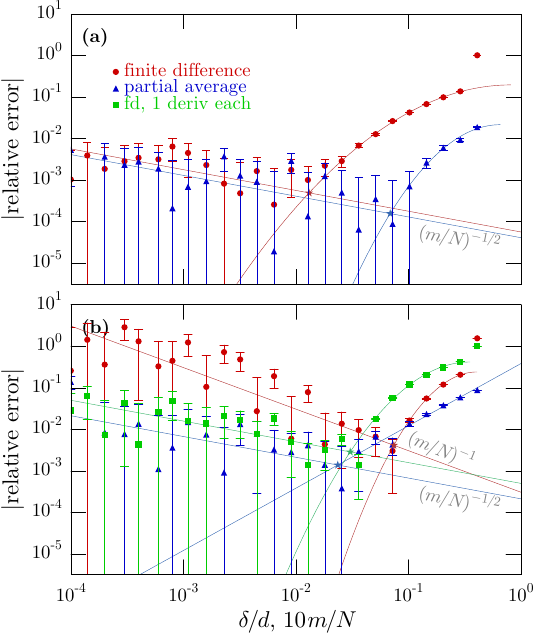}
  \end{center}
  \caption
  {
  Magnitude of the relative error in the computation of (a)
  one spatial derivative and (b) two derivatives 
  of the Casimir potential for 2 thin planes, 
  for susceptibility parameter $\hat{\chi}=1$.
  The plots compare the performance of the finite-difference
  (circles)
  and partial-averaging (triangles) methods; the second derivative
  additionally has one finite difference applied to each surface
  (squares).
  The finite-difference data are plotted with respect to the
  spatial displacement $\delta$, normalized to the
  atom--surface distance $d$, while the
  partial-averaging results are plotted as a function
  of the averaging fraction $m/N$, where $m$ appears
  in Eq.~(\ref{eq:sym-partial-avg}) and $N$ is the number
  of points per discrete path.
  The calculations employed paths with $N=10^5$ points per path,
  averaging over $10^{9}$ paths.
  Error bars delimit one standard error.
  The lines are fits to the data,
  and the intersection (marked by a star) gives the estimated best 
  performance of the method.
  Data points within each method are statistically independent.
    \label{fig:gradconv2plane-chi1}
  }
\end{figure}

\begin{figure}[tb]
  \begin{center}
\includegraphics[width=\columnwidth]{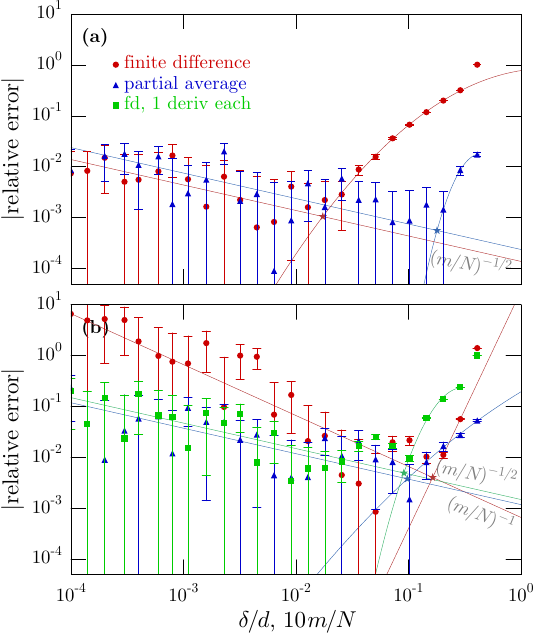}
  \end{center}
  \caption
  {
  Magnitude of the relative error in the computation of (a)
  one spatial derivative and (b) two derivatives 
  of the Casimir potential for 2 thin planes, 
  for susceptibility parameter $\hat{\chi}=0.01$.
  Details are as in Fig.~\ref{fig:gradconv2plane-chi1}.
    \label{fig:gradconv2plane-chi0.01}
  }
\end{figure}

The numerical methods of Section~\ref{sec:forces} 
for computing derivatives for 
Casimir-energy path integrals
are similar to the partial-averaging
method applied to the Casimir--Polder path integral in the
previous section.  
Fig.~\ref{fig:upto2derivs} shows the results of such a computation,
calculating the Casimir force (first derivative of the Casimir energy) 
and the curvature (second derivative of the Casimir energy) 
between two thin dielectric plates,
where the total susceptibility and the relative permittivity are
\begin{equation}
  \chi(z) = \hat\chi \delta(z)+\hat\chi \delta(z-d),
  \qquad \epsilon_\mathrm{r}(z)=1+\chi(z).
  \label{deltatwochipot}
\end{equation}
Here, viewing the delta-function limit as an idealization of
a square profile of a dielectric of finite thickness $a$, then 
$\hat{\chi}=\chi a$, where $\chi$ is the susceptibility of the medium.
The plot compares the numerical data to the 
exact Casimir force or curvature. When normalized to the
potential in the perfect-conductor limit 
($E\subEM/A = -\pi^{2}\hbar c/720 d^3$),
the Casimir energy has the form
\begin{equation}
    \gamma\subTE(\chi)= -\frac{180}{\pi^4}\int_0^\infty\! d\xi\,\xi^{2}
    \int_1^\infty\! dp\,p \log\!\left(1- r^2e^{-p\xi}\right)
    \label{gammaTE}
\end{equation}
in terms of the Fresnel coefficient
$r = -\xi(\hat{\chi}/d)/(2p+\xi\hat{\chi}/d)$.
This expression depends on
$\hat\chi$ and $d$ only in the combination
$\hat\chi/d$;
in the limit $d\gg \hat\chi$, this efficiency is
$\gamma\subTE\approx{27 \hat\chi^2}/{4\pi^4d^2}$, for a $d^{-5}$ far-field distance dependence,
while in the limit $d\ll\hat\chi$, the efficiency is
$\gamma\subTE\approx 1/2$ (i.e.,
the strong-coupling limit).
The data are compared to derivatives of this expression
with respect to $d$.

The numerical method solves the exponential form 
(\ref{eq:casimir_energy_exp})
of the path integral; the only error that is not statistical
is in treating the two surfaces as independent. That is, the
error is in treating the occupation time for each delta-function surface
independently, not both surfaces together.
This explains the much better performance of the finite-difference method
compared to  Figs.~\ref{fig:convergence} and \ref{fig:convergence-chi-inf}.
The numerical tests implement Eqs.~(\ref{expform1deriv})
and (\ref{expform2deriv}), as appropriate for an exponential
path integral.
The discrete paths over $N$ points are averaged over all sub-paths between
the points $x_{j+1}$ and $x_j$; in the one-body approximation, the 
appropriate quantity is the local time (rather, its generating function)
and its derivative, detailed in
Appendix~\ref{app:localtime}.
Values for $\cT$ and $s$ are Monte-Carlo sampled. We also take advantage
of the fact that any contributing path must cross any plane between
the surfaces.  That is, we may start paths midway between the two planes,
with a reweighting factor of
$N(x_+-x_-)/2$,
where $x_{+}$ is the smallest path coordinate such that $x_+>x_0$
(or just $x_0$ if no other such point exists),
and $x_{-}$ is the largest path coordinate such that $x_-<x_0$
(or just $x_0$ if no other such point exists).

The partial-averaging method amounts to looking for path 
segments that get close enough to the boundaries; details
are given in Sec.~\ref{sec:force}, particularly 
Eqs.~(\ref{expform1deriv}), (\ref{eq:partialavgerr2plane}), and
(\ref{expform2deriv}).
Similar to the Casimir--Polder case, 
renormalization of the one-body energy is necessary to obtain
a finite result.
For the first derivative, we compute the one-body energy
for the undifferentiated surface, and subtract it from the total
energy.
For the second derivative (\ref{expform2deriv}), where
one derivative is applied to each surface, no renormalization is
necessary.

For the finite-difference method, with renormalization
of the one-body energies, the path integral has the
form
\begin{equation}
  I(\mathbf{R}_1,\mathbf{R}_2) = p(\mathbf{R}_1,\mathbf{R}_2) - p_1(\mathbf{R}_1) - p_2(\mathbf{R}_2),
\end{equation}
where $p$ is the two-body energy, $p_j$ is the one-body energy of 
body $j$, and $\mathbf{R}_j$ is the position of body $j$.
The first derivative is computed according to a
standard, second-order finite difference, which
for the $z$ component reads
\begin{align}
  \hat{z}\cdot\nabla_{\mathbf{R}_1} I &=  
  \frac{1}{2\delta}
  \Big[p(\mathbf{R}_1+\hat{z}\delta,\mathbf{R}_2) - p_1(\mathbf{R}_1+\hat{z}\delta)  
  \\\nonumber
  &\hspace{15mm}-p(\mathbf{R}_1-\hat{z}\delta,\mathbf{R}_2) + p_1(\mathbf{R}_1-\hat{z}\delta)\Big],
\end{align}
where $\delta$ is some small perturbation.
The second derivative is also computed according to a second-order
finite difference,
\begin{align}\nonumber
  (\hat{z}\cdot\nabla_{\mathbf{R}_1})^2 I &=  
  \frac{1}{\delta^2}
  \Big[p(\mathbf{R}_1+\hat{z}\delta,\mathbf{R}_2)
  - p_1(\mathbf{R}_1+\hat{z}\delta)  
  \\\nonumber
  &\hspace{12mm}+p(\mathbf{R}_1-\hat{z}\delta,\mathbf{R}_2) - p_1(\mathbf{R}_1-\hat{z}\delta)
  \\
  &\hspace{12mm}-2p(\mathbf{R}_1,\mathbf{R}_2) + 2p_1(\mathbf{R}_1)\Big]
 .
\end{align}
A better comparison with the partially averaged second derivative
is 
\begin{align}
  (\hat{z}\cdot\nabla_{\mathbf{R}_1})^2 I 
  &= - \frac{1}{4\delta^2}
  \Big[p(\mathbf{R}_1\!+\!\hat{z}\delta,\mathbf{R}_2\!+\!\hat{z}\delta) 
  \nonumber\\
  &\hspace{15mm}  -p(\mathbf{R}_1\!  -\!\hat{z}\delta,\mathbf{R}_2\!+\!\hat{z}\delta) 
  \nonumber\\
  &\hspace{15mm}-p(\mathbf{R}_1\!+\!\hat{z}\delta,\mathbf{R}_2\!-\!\hat{z}\delta)  
  \nonumber\\
  &\hspace{15mm}+p(\mathbf{R}_1\!-\!\hat{z}\delta,\mathbf{R}_2\!-\!\hat{z}\delta)\Big],
\end{align}
applying one derivative to each interface.

We observe good agreement with the exact results
(Fig.~\ref{fig:upto2derivs}).
The calculations employed $10^9$ paths, each of $10^5$ points per path.
The first derivative used partial averaging over 300 points,
while the second derivative used 250 points.
The relative error in the first derivative ranges from 
$3.5\times 10^{-3}$ for small $\hat{\chi}$ to 
$4\times 10^{-4}$ for large $\hat{\chi}$, and
for the second derivative, the relative error ranges
from $1.2\%$ at the low end to around $3\times 10^{-3}$
at $\hat{\chi}=1$, though the error rises to about $4\%$ for the
highest $\hat{\chi}$.

Figs.~\ref{fig:gradconv2plane-chi1} and \ref{fig:gradconv2plane-chi0.01}
show the errors of the partial-average method vs.\ the finite difference
(and in the case of the second derivative, of one finite difference applied
to each side).
These plots display the relative error vs.\ the normalized separation
$\delta/d$ for the 
finite-difference methods, or the partial-averaging fraction $m/N$, 
as described in Section~\ref{sec:force}.
As in Figs.~\ref{fig:convergence} and \ref{fig:convergence-chi-inf},
the lines for small $\hat{\chi}$ are curve fits to
$a(\delta/d)^{-1/2}$ or $a(10m/N)^{-1/2}$ in 
Figs.~\ref{fig:gradconv2plane-chi1}a and \ref{fig:gradconv2plane-chi0.01}a,
and to 
$a(\delta/d)^{-1/2}$, $a(10m/N)^{-1/2}$, or
$a(\delta/d)^{-1}$, as indicated in
Figs.~\ref{fig:gradconv2plane-chi1}b and \ref{fig:gradconv2plane-chi0.01}b
(fit ranges indicated in App.~\ref{app:endpoints}, Table~\ref{table:table2}).
For large $\hat{\chi}$, the curve fits are second-order polynomials to the log of the data
(on both axes).
The intersection of the small-$\hat{\chi}$ and large-$\hat{\chi}$ curves
is marked with a star, and gives an estimate of the best error of the method.
The partial-averaging method outperforms the finite-difference method
for the first derivative in Fig.~\ref{fig:gradconv2plane-chi1}a, for $\hat{\chi}=1$
(where the best error is $1.5\times 10^{-4}$ for partial averaging, vs.\ 
$4.9\times 10^{-4}$ for finite difference).  
The partial average still outperforms the finite difference for the
second derivative in Fig.~\ref{fig:gradconv2plane-chi1}b, again for $\hat{\chi}=1$
(where the best error is $1.4\times 10^{-3}$ for partial averaging, vs.\ 
$4.1\times 10^{-3}$ for finite difference, or $2.8\times 10^{-3}$ for finite difference,
one derivative applied to each side).  
For $\hat{\chi}=0.01$ (Fig.~\ref{fig:gradconv2plane-chi0.01}), 
while the advantage persists for one derivative
(where the best error is $5.6\times 10^{-4}$ for partial averaging, vs.\ 
$1.0\times 10^{-3}$ for finite difference);
however, for the second derivative, the performance is roughly the same
($3.8\times 10^{-3}$ for partial averaging, vs.\ 
$4.0\times 10^{-3}$ for finite difference, or $4.9\times 10^{-3}$ for finite difference,
one derivative applied to each side).  
For $\hat{\chi}=100$ and the $\hat{\chi}\longrightarrow\infty$ limit (both not shown), 
the methods are comparable in error for both the first
and second derivative.
So the partial average sometimes performs substantially better than 
finite differences, and never performs substantially worse.
But it should also be appreciated that this partial average method is much
faster than finite differences near the optimum value, because of the 
large time jumps near the surfaces involved.

The method that we have tested here clearly extends to a general geometry.
It involves counting the intersections with one (or both) surfaces, and replacing one intersection 
(or one for each surface) with its derivative.  It is thus not
geometry-specific.
The only requirement for partial averaging is that
the typical distance covered during a large jump is small 
compared to a length scale where the nearest surface can 
be considered flat
(assuming that a flat-surface approximation is employed, for example,
in a sphere-sphere geometry \cite{Schoger2022}).

\section{Summary}

We have developed methods for computing derivatives 
of Casimir--Polder and Casimir-type path integrals.  
Such derivatives are notable for being difficult to 
compute using finite differences, because they converge 
badly---a path only contributes if it marginally touches a surface,
but has a large contribution, leading to a large sample variance.
For Casimir--Polder-type path integrals, our method involves
partial averaging over the region around the atom, which cuts off
the sample variance from diverging.
For Casimir-type integrals, the method involves counting all 
intersections with surfaces, 
and differentiating one of the subpaths.  The partial averaging method
there involves simply skipping large steps in time, leading to a 
large gain in computation time.
In a future publication, we will apply this method to 
the Casimir--Polder potential for transverse-magnetic modes, 
which involves a second derivative.

\appendix
\section{Brownian-Bridge Generation}
\label{app:openbridge}

The numerical evaluation of the path integrals in this paper
and other work on worldline methods
involves the generation of discrete, stochastic paths.
The basic stochastic process is the Brownian bridge $B(t)$,
which in its most basic form is subject to 
initial and final conditions $B(0)=B(1)=0$, and
diffuses continuously as a Wiener process for intermediate times
[i.e.,
$\dlangle dB(t)\drangle = 0$ and
$\dlangle dB(t)\,dB(t')\drangle= \delta(t-t')\,dt$
for $t,t'\in(0,1)$].
The discrete representation of a Brownian bridge
follows from the more general probability density (\ref{eq:pathmeasure}).
Samples $B_j$ of a Brownian bridge in $N$ discrete steps may
then be generated in a numerical calculation via the recurrence relation
\begin{equation}
  B_j = \sqrt{\dfrac{c_j}{N}}\,z_j + c_j B_{j-1}\quad (j = 1,\ldots,N-1),
  \label{bridgerecur}
\end{equation}
where $B_0=B_N=0$, the $z_k$ are random deviates sampled from 
the standard-normal distribution, and the recursion coefficients are
\begin{equation}
    c_j :=\dfrac{N-j}{N-j+1} .
  \label{bridgerecurcoef}
\end{equation}
This algorithm was discussed in previous work~\cite{Mackrory2016}
(and is a variation of the ``v-loop'' algorithm of 
Gies~\etal~\cite{Gies2003}).

In the numerical methods of this paper, it is necessary
to generate samples of more general, ``open'' Brownian bridges
that are subject to different initial and final conditions.
(Specifically, this applies to the partial-averaging method
as discussed in Sec.~\ref{sec:partial_averaging_numerical}, and
the finer path sampling around the ``extremal'' path point
in the same section.)
Under the conditions $B(0)=a$ and $B(t)=b$, the above recurrence 
generalizes to
\begin{align}
  B_j =& \sqrt{c_j\,\Delta t}\,z_j + c_j (B_{j-1}-b)+b \nonumber\\
    &\hspace{3cm} (j = 1,\ldots,N-1),
  \label{bridgerecurgen}
\end{align}
where $B_0=a$, $B_N=b$,
the bridge takes $N$ intermediate steps (each of time $\Delta t=t/N$),
and the $c_j$ are still given by Eq.~(\ref{bridgerecurcoef}).

In the partial-averaging construction of paths in 
Sec.~\ref{sec:partial_average},
some portion of a path is constructed according to some consideration,
and the generalized bridge recursion serves to fill in the rest
of the path. For example, 
sampling from the Hermite--Gaussian density (\ref{eq:Hmstep}),
after partial averaging, gives the first $m$ path steps
at the beginning and end of the path, and a Brownian
bridge must then complete the path in $N-2m$ steps.
To write out the algorithm more explicitly for this problem,
suppose the ``whole'' path corresponds to an $N$-step, discrete
Brownian bridge $B(t)$ with $B(1)=B(0)=0$, and that 
we take the path coordinate
$B_k=B(t=k/N)$
after $k$ steps to be given.
The above recursion then constructs
a bridge from $B_k$ to $B_0$ in $N_k:=N-k$ steps over a running time
$1-(k/N) = N_k/N$.  
Because there are $N_k$ time steps of duration
$\Delta t=(N_k/N)/N_k=1/N$,
the recurrence (\ref{bridgerecurgen}) reads
\begin{align}
  B_j =& \sqrt{\dfrac{c_j}{N}}\,z_j + c_j (B_{j-1}-b)+b \nonumber\\
    &\hspace{3cm} (j = k+1,\ldots,N-1)
  \label{bridgerecurgenk}
\end{align}
when adapted to this situation.  
The coefficients $c_j=(N_k-j)/(N_k-j+1)$ reduce to the expression
(\ref{bridgerecurcoef}) in terms of $N$
if the index $j$ is shifted by $k$,
so that $j$ starts at $k+1$ [note that this shift has already been
applied in the recursion (\ref{bridgerecurgenk})].
Thus, the recursion (\ref{bridgerecurgenk}) has 
the form of the \textit{last} $N-k$ steps
of the recursion (\ref{bridgerecurgen})---an intuitive
way to think of ``completing'' the path once the first $k$
steps are fixed.

\vspace{4ex}
\section{Local Time}\label{app:localtime}

For a stochastic process $y(t)$, the local time is defined as \cite{Borodin2002}
\begin{equation}
  \ell[y(t);d] := \int_0^t d\tau \delta [y(\tau)-d]. 
\end{equation}
The local time quantifies the amount of time that the process spends near a boundary at $d$ during the interval $[0,t]$. It is related to the sojourn time $T_s$, which measures the amount of time that the process spends in the state $y(t)>d$, through the derivative 
\begin{equation}
  \ell[y(t);d] = -\partial_d T_s[y(t);d].
\end{equation}
(See Ref.~\cite{Mackrory2016} for a definition and some statistics
of the sojourn time.)
The local times of a Wiener process with 
$y(0)=a$ and $y(t)=b$ are distributed according to the probability density
\begin{align}
 f_\ell (x) 
 &= [1-e^{[(b-a)^2-(|a-d|+|b-d|)^2]/2t}] \,\delta(x-0^+)\nonumber\\
 &+\frac{1}{t}(x+|a-d|+|b-d|)\nonumber\\
 &\hspace{5mm}\times e^{[(b-a)^2-(x+|a-d|+|b-d|)^2]/2t}.
\end{align}
An important quantity for the local time that we use in the simulations is the moment generating function, which takes the form 
\begin{align}
  \Bigdlangle e^{-s\ell} \Bigdrangle&
    =1-\sqrt{\frac{\pi t}{2}}\, s\, \nonumber\\
  &\times  \exp\left[\frac{(b-a)^2+s^2 t^2+2st(|b-d|+|a-d|)}{2t}\right] \nonumber\\
  &\times \text{erfc}\left(\frac{| b-d| +|a- d| +s t}{\sqrt{2t}}\right) ,
\end{align}
again with $y(0)=a$, $y(t)=b$.
This expression, along with its derivative with respect to $d$, is used
in the simulations in Sec.~\ref{sec:casforcecurve}.

\section{Endpoints for Curve Fits}\label{app:endpoints}

Here the endpoints for the curve fits are listed for
Figs.~\ref{fig:convergence} and \ref{fig:convergence-chi-inf} 
(Table~\ref{table:table1}), and
\ref{fig:gradconv2plane-chi1} and \ref{fig:gradconv2plane-chi0.01}
(Table~\ref{table:table2}).
We list these since the fit results may depend somewhat on the
choice of endpoint.
In many cases, points near the crossover between the two curves
are excluded because it is not clear where they should be assigned.

\begin{table*}[t]
\begin{tabular}{|c|c|c|c|c|c|}\hline
\multirow{2}{*}{\textbf{Figure}}&\multirow{2}{*}{\textbf{curve}}&\multicolumn{2}{c|}{\textbf{small} $\boldsymbol{\chi}$}& \multicolumn{2}{c|}{\textbf{large} $\boldsymbol{\chi}$}\\\cline{3-6}
 & & \textbf{low} & \textbf{high}& \textbf{low} & \textbf{high}\\\hline
\ref{fig:convergence}a & finite difference $(\delta/d)$ & $10^{-4}$&$4.5\times 10^{-3}$&$9.1\times 10^{-3}$&$4.096\times 10^{-1}$\\\hline
\ref{fig:convergence}a & partial average $(m/N)$ & $10^{-4}$&$4.5\times 10^{-3}$&$3.62\times 10^{-2}$&$4.096\times 10^{-1}$\\\hline
\ref{fig:convergence}b & partial average $(m/N)$ & $10^{-4}$&$1.81\times 10^{-2}$&$3.62\times 10^{-2}$&$4.096\times 10^{-1}$\\\hline
\ref{fig:convergence-chi-inf}a & finite difference $(\delta/d)$ & $10^{-4}$&$1.1\times 10^{-3}$&$3.2\times 10^{-3}$&$7.24\times 10^{-2}$\\\hline
\ref{fig:convergence-chi-inf}a & partial average $(m/N)$ & $10^{-4}$&$1.81\times 10^{-2}$&$3.62\times 10^{-2}$&$4.096\times 10^{-1}$\\\hline
\ref{fig:convergence-chi-inf}b & partial average $(m/N)$ & $10^{-4}$&$2.56\times 10^{-2}$&$5.12\times 10^{-2}$&$4.096\times 10^{-1}$\\\hline
\end{tabular}
  \caption
  {
  Fit ranges for Figs.~\ref{fig:convergence} and \ref{fig:convergence-chi-inf}.
    \label{table:table1}
  }
\end{table*}

\begin{table*}[t]
\begin{tabular}{|c|c|c|c|c|c|}\hline
\multirow{2}{*}{\textbf{Figure}}&\multirow{2}{*}{\textbf{curve}}&\multicolumn{2}{c|}{\textbf{small} $\boldsymbol{\hat{\chi}}$}& \multicolumn{2}{c|}{\textbf{large} $\boldsymbol{\hat{\chi}}$}\\\cline{3-6}
 & & \textbf{low} & \textbf{high}& \textbf{low} & \textbf{high}\\\hline
\ref{fig:gradconv2plane-chi1}a & finite difference $(\delta/d)$ & $10^{-4}$&$9.1\times 10^{-3}$&$2.56\times 10^{-2}$&$2.897\times 10^{-1}$\\\hline
\ref{fig:gradconv2plane-chi1}a & partial average $(10m/N)$ & $10^{-4}$&$5.12\times 10^{-2}$&$1.024\times 10^{-1}$&$4.096\times 10^{-1}$\\\hline
\ref{fig:gradconv2plane-chi1}b & finite difference $(\delta/d)$ & $10^{-4}$&$5.12\times 10^{-2}$&$1.024\times 10^{-1}$&$2.897\times 10^{-1}$\\\hline
\ref{fig:gradconv2plane-chi1}b & partial average $(10m/N)$ & $10^{-4}$&$2.56\times 10^{-2}$&$5.12\times 10^{-2}$&$4.096\times 10^{-1}$\\\hline
\ref{fig:gradconv2plane-chi1}b & fd, 1 derivative each $(\delta/d)$ & $10^{-4}$&$2.56\times 10^{-2}$&$5.12\times 10^{-2}$&$2.897\times 10^{-1}$\\\hline
\ref{fig:gradconv2plane-chi0.01}a & finite difference $(\delta/d)$ & $10^{-4}$&$1.28\times 10^{-2}$&$2.56\times 10^{-2}$&$2.897\times 10^{-1}$\\\hline
\ref{fig:gradconv2plane-chi0.01}a & partial average $(10m/N)$ & $10^{-4}$&$1.449\times 10^{-1}$&$2.048\times 10^{-1}$&$4.096\times 10^{-1}$\\\hline
\ref{fig:gradconv2plane-chi0.01}b & finite difference $(\delta/d)$ & $10^{-4}$&$1.024\times 10^{-1}$&$2.048\times 10^{-1}$&$2.897\times 10^{-1}$\\\hline
\ref{fig:gradconv2plane-chi0.01}b & partial average $(10m/N)$ & $10^{-4}$&$7.24\times 10^{-2}$&$1.449\times 10^{-1}$&$4.096\times 10^{-1}$\\\hline
\ref{fig:gradconv2plane-chi0.01}b & fd, 1 derivative each $(\delta/d)$ & $10^{-4}$&$7.24\times 10^{-2}$&$1.024\times 10^{-1}$&$2.897\times 10^{-1}$\\\hline
\end{tabular}
  \caption
  {
  Fit ranges for Figs.~\ref{fig:gradconv2plane-chi1} and
  \ref{fig:gradconv2plane-chi0.01}.
    \label{table:table2}
  }
\end{table*}

\bibliographystyle{aipnum4-2}
%

\begin{thebibliography}{36}%
\makeatletter
\providecommand \@ifxundefined [1]{%
 \@ifx{#1\undefined}
}%
\providecommand \@ifnum [1]{%
 \ifnum #1\expandafter \@firstoftwo
 \else \expandafter \@secondoftwo
 \fi
}%
\providecommand \@ifx [1]{%
 \ifx #1\expandafter \@firstoftwo
 \else \expandafter \@secondoftwo
 \fi
}%
\providecommand \natexlab [1]{#1}%
\providecommand \enquote  [1]{``#1''}%
\providecommand \bibnamefont  [1]{#1}%
\providecommand \bibfnamefont [1]{#1}%
\providecommand \citenamefont [1]{#1}%
\providecommand \href@noop [0]{\@secondoftwo}%
\providecommand \href [0]{\begingroup \@sanitize@url \@href}%
\providecommand \@href[1]{\@@startlink{#1}\@@href}%
\providecommand \@@href[1]{\endgroup#1\@@endlink}%
\providecommand \@sanitize@url [0]{\catcode `\\12\catcode `\$12\catcode
  `\&12\catcode `\#12\catcode `\^12\catcode `\_12\catcode `\%12\relax}%
\providecommand \@@startlink[1]{}%
\providecommand \@@endlink[0]{}%
\providecommand \url  [0]{\begingroup\@sanitize@url \@url }%
\providecommand \@url [1]{\endgroup\@href {#1}{\urlprefix }}%
\providecommand \urlprefix  [0]{URL }%
\providecommand \Eprint [0]{\href }%
\providecommand \doibase [0]{https://doi.org/}%
\providecommand \selectlanguage [0]{\@gobble}%
\providecommand \bibinfo  [0]{\@secondoftwo}%
\providecommand \bibfield  [0]{\@secondoftwo}%
\providecommand \translation [1]{[#1]}%
\providecommand \BibitemOpen [0]{}%
\providecommand \bibitemStop [0]{}%
\providecommand \bibitemNoStop [0]{.\EOS\space}%
\providecommand \EOS [0]{\spacefactor3000\relax}%
\providecommand \BibitemShut  [1]{\csname bibitem#1\endcsname}%
\let\auto@bib@innerbib\@empty
\bibitem [{\citenamefont {Casimir}(1948)}]{Casimir1948a}%
  \BibitemOpen
  \bibfield  {author} {\bibinfo {author} {\bibfnamefont {H.~B.~G.}\
  \bibnamefont {Casimir}},\ }\href@noop {} {\bibfield  {journal} {\bibinfo
  {journal} {{Proc.\ K.\ Ned.\ Akad.\ Wet.}}\ }\textbf {\bibinfo {volume}
  {51}},\ \bibinfo {pages} {793} (\bibinfo {year} {1948})}\BibitemShut
  {NoStop}%
\bibitem [{\citenamefont {Casimir}\ and\ \citenamefont
  {Polder}(1948)}]{CasimirPolder1948}%
  \BibitemOpen
  \bibfield  {author} {\bibinfo {author} {\bibfnamefont {H.~B.~G.}\
  \bibnamefont {Casimir}}\ and\ \bibinfo {author} {\bibfnamefont
  {D.}~\bibnamefont {Polder}},\ }\href@noop {} {\bibfield  {journal} {\bibinfo
  {journal} {{Phys.\ Rev.}}\ }\textbf {\bibinfo {volume} {73}},\ \bibinfo
  {pages} {360} (\bibinfo {year} {1948})}\BibitemShut {NoStop}%
\bibitem [{\citenamefont {Lifshitz}(1956)}]{Lifshitz1956}%
  \BibitemOpen
  \bibfield  {author} {\bibinfo {author} {\bibfnamefont {E.}~\bibnamefont
  {Lifshitz}},\ }\href@noop {} {\bibfield  {journal} {\bibinfo  {journal} {{J.\
  Exper.\ Theoret.\ Phys.\ USSR}}\ }\textbf {\bibinfo {volume} {29}},\ \bibinfo
  {pages} {94} (\bibinfo {year} {1956})}\BibitemShut {NoStop}%
\bibitem [{\citenamefont {Chan}\ \emph {et~al.}(2001)\citenamefont {Chan},
  \citenamefont {Aksyuk}, \citenamefont {Kleiman}, \citenamefont {Bishop},\
  and\ \citenamefont {Capasso}}]{Chan2001b}%
  \BibitemOpen
  \bibfield  {author} {\bibinfo {author} {\bibfnamefont {H.~B.}\ \bibnamefont
  {Chan}}, \bibinfo {author} {\bibfnamefont {V.~A.}\ \bibnamefont {Aksyuk}},
  \bibinfo {author} {\bibfnamefont {R.~N.}\ \bibnamefont {Kleiman}}, \bibinfo
  {author} {\bibfnamefont {D.~J.}\ \bibnamefont {Bishop}},\ and\ \bibinfo
  {author} {\bibfnamefont {F.}~\bibnamefont {Capasso}},\ }\href@noop {}
  {\bibfield  {journal} {\bibinfo  {journal} {Science}\ }\textbf {\bibinfo
  {volume} {291}},\ \bibinfo {pages} {1941} (\bibinfo {year}
  {2001})}\BibitemShut {NoStop}%
\bibitem [{\citenamefont {Harber}\ \emph {et~al.}(2005)\citenamefont {Harber},
  \citenamefont {Obrecht}, \citenamefont {McGuirk},\ and\ \citenamefont
  {Cornell}}]{Harber2005}%
  \BibitemOpen
  \bibfield  {author} {\bibinfo {author} {\bibfnamefont {D.~M.}\ \bibnamefont
  {Harber}}, \bibinfo {author} {\bibfnamefont {J.~M.}\ \bibnamefont {Obrecht}},
  \bibinfo {author} {\bibfnamefont {J.~M.}\ \bibnamefont {McGuirk}},\ and\
  \bibinfo {author} {\bibfnamefont {E.~A.}\ \bibnamefont {Cornell}},\
  }\href@noop {} {\bibfield  {journal} {\bibinfo  {journal} {Phys. Rev. A}\
  }\textbf {\bibinfo {volume} {72}},\ \bibinfo {pages} {033610} (\bibinfo
  {year} {2005})}\BibitemShut {NoStop}%
\bibitem [{\citenamefont {Gies}, \citenamefont {Langfeld},\ and\ \citenamefont
  {Moyaerts}(2003)}]{Gies2003}%
  \BibitemOpen
  \bibfield  {author} {\bibinfo {author} {\bibfnamefont {H.}~\bibnamefont
  {Gies}}, \bibinfo {author} {\bibfnamefont {K.}~\bibnamefont {Langfeld}},\
  and\ \bibinfo {author} {\bibfnamefont {L.}~\bibnamefont {Moyaerts}},\
  }\href@noop {} {\bibfield  {journal} {\bibinfo  {journal} {{J.\ High Energy
  Phys.}}\ }\textbf {\bibinfo {volume} {2003}},\ \bibinfo {pages} {018}
  (\bibinfo {year} {2003})}\BibitemShut {NoStop}%
\bibitem [{\citenamefont {Gies}\ and\ \citenamefont
  {Klingm\"{u}ller}(2006)}]{Gies2006a}%
  \BibitemOpen
  \bibfield  {author} {\bibinfo {author} {\bibfnamefont {H.}~\bibnamefont
  {Gies}}\ and\ \bibinfo {author} {\bibfnamefont {K.}~\bibnamefont
  {Klingm\"{u}ller}},\ }\href@noop {} {\bibfield  {journal} {\bibinfo
  {journal} {{J.\ Phys.\ A}}\ }\textbf {\bibinfo {volume} {39}},\ \bibinfo
  {pages} {6415} (\bibinfo {year} {2006})}\BibitemShut {NoStop}%
\bibitem [{\citenamefont {Gies}\ and\ \citenamefont
  {Klingm{\"u}ller}(2006{\natexlab{a}})}]{Gies2006b}%
  \BibitemOpen
  \bibfield  {author} {\bibinfo {author} {\bibfnamefont {H.}~\bibnamefont
  {Gies}}\ and\ \bibinfo {author} {\bibfnamefont {K.}~\bibnamefont
  {Klingm{\"u}ller}},\ }\href@noop {} {\bibfield  {journal} {\bibinfo
  {journal} {{Phys.\ Rev.\ Lett.}}\ }\textbf {\bibinfo {volume} {96}},\
  \bibinfo {pages} {220401} (\bibinfo {year} {2006}{\natexlab{a}})}\BibitemShut
  {NoStop}%
\bibitem [{\citenamefont {Gies}\ and\ \citenamefont
  {Klingm{\"u}ller}(2006{\natexlab{b}})}]{Gies2006c}%
  \BibitemOpen
  \bibfield  {author} {\bibinfo {author} {\bibfnamefont {H.}~\bibnamefont
  {Gies}}\ and\ \bibinfo {author} {\bibfnamefont {K.}~\bibnamefont
  {Klingm{\"u}ller}},\ }\href@noop {} {\bibfield  {journal} {\bibinfo
  {journal} {{Phys.\ Rev.\ Lett.}}\ }\textbf {\bibinfo {volume} {97}},\
  \bibinfo {pages} {220405} (\bibinfo {year} {2006}{\natexlab{b}})}\BibitemShut
  {NoStop}%
\bibitem [{\citenamefont {Gies}\ and\ \citenamefont
  {Klingm{\"u}ller}(2006{\natexlab{c}})}]{Gies2006d}%
  \BibitemOpen
  \bibfield  {author} {\bibinfo {author} {\bibfnamefont {H.}~\bibnamefont
  {Gies}}\ and\ \bibinfo {author} {\bibfnamefont {K.}~\bibnamefont
  {Klingm{\"u}ller}},\ }\href@noop {} {\bibfield  {journal} {\bibinfo
  {journal} {{Phys.\ Rev.\ D}}\ }\textbf {\bibinfo {volume} {74}},\ \bibinfo
  {pages} {045002} (\bibinfo {year} {2006}{\natexlab{c}})}\BibitemShut
  {NoStop}%
\bibitem [{\citenamefont {Klingm{\"u}ller}\ and\ \citenamefont
  {Gies}(2008)}]{Klingmuller2008}%
  \BibitemOpen
  \bibfield  {author} {\bibinfo {author} {\bibfnamefont {K.}~\bibnamefont
  {Klingm{\"u}ller}}\ and\ \bibinfo {author} {\bibfnamefont {H.}~\bibnamefont
  {Gies}},\ }\href@noop {} {\bibfield  {journal} {\bibinfo  {journal} {{J.\
  Phys.\ A}}\ }\textbf {\bibinfo {volume} {41}},\ \bibinfo {pages} {164042}
  (\bibinfo {year} {2008})}\BibitemShut {NoStop}%
\bibitem [{\citenamefont {Weber}\ and\ \citenamefont {Gies}(2009)}]{Weber2009}%
  \BibitemOpen
  \bibfield  {author} {\bibinfo {author} {\bibfnamefont {A.}~\bibnamefont
  {Weber}}\ and\ \bibinfo {author} {\bibfnamefont {H.}~\bibnamefont {Gies}},\
  }\href@noop {} {\bibfield  {journal} {\bibinfo  {journal} {{Phys.\ Rev.\
  D.}}\ }\textbf {\bibinfo {volume} {80}},\ \bibinfo {pages} {065033} (\bibinfo
  {year} {2009})}\BibitemShut {NoStop}%
\bibitem [{\citenamefont {Fosco}, \citenamefont {Lombardo},\ and\ \citenamefont
  {Mazzitelli}(2010)}]{Fosco2010}%
  \BibitemOpen
  \bibfield  {author} {\bibinfo {author} {\bibfnamefont {C.}~\bibnamefont
  {Fosco}}, \bibinfo {author} {\bibfnamefont {F.}~\bibnamefont {Lombardo}},\
  and\ \bibinfo {author} {\bibfnamefont {F.}~\bibnamefont {Mazzitelli}},\
  }\href@noop {} {\bibfield  {journal} {\bibinfo  {journal} {{Phys.\ Lett.\
  B}}\ }\textbf {\bibinfo {volume} {690}},\ \bibinfo {pages} {189} (\bibinfo
  {year} {2010})}\BibitemShut {NoStop}%
\bibitem [{\citenamefont {Weber}\ and\ \citenamefont {Gies}(2010)}]{Weber2010}%
  \BibitemOpen
  \bibfield  {author} {\bibinfo {author} {\bibfnamefont {A.}~\bibnamefont
  {Weber}}\ and\ \bibinfo {author} {\bibfnamefont {H.}~\bibnamefont {Gies}},\
  }\href@noop {} {\bibfield  {journal} {\bibinfo  {journal} {{Phys.\ Rev.\
  Lett.}}\ }\textbf {\bibinfo {volume} {105}},\ \bibinfo {pages} {040403}
  (\bibinfo {year} {2010})}\BibitemShut {NoStop}%
\bibitem [{\citenamefont {Aehlig}\ \emph {et~al.}(2011)\citenamefont {Aehlig},
  \citenamefont {Dietert}, \citenamefont {Fischbacher},\ and\ \citenamefont
  {Gerhard}}]{Aehlig2011}%
  \BibitemOpen
  \bibfield  {author} {\bibinfo {author} {\bibfnamefont {K.}~\bibnamefont
  {Aehlig}}, \bibinfo {author} {\bibfnamefont {H.}~\bibnamefont {Dietert}},
  \bibinfo {author} {\bibfnamefont {T.}~\bibnamefont {Fischbacher}},\ and\
  \bibinfo {author} {\bibfnamefont {J.}~\bibnamefont {Gerhard}},\ }\href@noop
  {} {} (\bibinfo {year} {2011}),\ \Eprint {https://arxiv.org/abs/1110.5936}
  {arXiv:1110.5936} \BibitemShut {NoStop}%
\bibitem [{\citenamefont {Sch{\"a}fer}, \citenamefont {Huet},\ and\
  \citenamefont {Gies}(2012)}]{Schafer2012}%
  \BibitemOpen
  \bibfield  {author} {\bibinfo {author} {\bibfnamefont {M.}~\bibnamefont
  {Sch{\"a}fer}}, \bibinfo {author} {\bibfnamefont {I.}~\bibnamefont {Huet}},\
  and\ \bibinfo {author} {\bibfnamefont {H.}~\bibnamefont {Gies}},\ }\href@noop
  {} {\bibfield  {journal} {\bibinfo  {journal} {{Int.\ J.\ Mod.\ Phys.\ Conf.\
  Ser.}}\ }\textbf {\bibinfo {volume} {14}},\ \bibinfo {pages} {511} (\bibinfo
  {year} {2012})}\BibitemShut {NoStop}%
\bibitem [{\citenamefont {Mazur}\ and\ \citenamefont {Heyl}(2015)}]{Mazur2015}%
  \BibitemOpen
  \bibfield  {author} {\bibinfo {author} {\bibfnamefont {D.}~\bibnamefont
  {Mazur}}\ and\ \bibinfo {author} {\bibfnamefont {J.~S.}\ \bibnamefont
  {Heyl}},\ }\href@noop {} {\bibfield  {journal} {\bibinfo  {journal} {\prd}\
  }\textbf {\bibinfo {volume} {91}},\ \bibinfo {pages} {065019} (\bibinfo
  {year} {2015})}\BibitemShut {NoStop}%
\bibitem [{\citenamefont {Sch{\"a}fer}, \citenamefont {Huet},\ and\
  \citenamefont {Gies}(2016)}]{Schafer2016}%
  \BibitemOpen
  \bibfield  {author} {\bibinfo {author} {\bibfnamefont {M.}~\bibnamefont
  {Sch{\"a}fer}}, \bibinfo {author} {\bibfnamefont {I.}~\bibnamefont {Huet}},\
  and\ \bibinfo {author} {\bibfnamefont {H.}~\bibnamefont {Gies}},\ }\href@noop
  {} {\bibfield  {journal} {\bibinfo  {journal} {J.\ Phys.\ A}\ }\textbf
  {\bibinfo {volume} {49}},\ \bibinfo {pages} {135402} (\bibinfo {year}
  {2016})}\BibitemShut {NoStop}%
\bibitem [{\citenamefont {Mackrory}, \citenamefont {Bhattacharya},\ and\
  \citenamefont {Steck}(2016)}]{Mackrory2016}%
  \BibitemOpen
  \bibfield  {author} {\bibinfo {author} {\bibfnamefont {J.~B.}\ \bibnamefont
  {Mackrory}}, \bibinfo {author} {\bibfnamefont {T.}~\bibnamefont
  {Bhattacharya}},\ and\ \bibinfo {author} {\bibfnamefont {D.~A.}\ \bibnamefont
  {Steck}},\ }\href@noop {} {\bibfield  {journal} {\bibinfo  {journal} {Phys.
  Rev. A}\ }\textbf {\bibinfo {volume} {94}},\ \bibinfo {pages} {042508}
  (\bibinfo {year} {2016})}\BibitemShut {NoStop}%
\bibitem [{\citenamefont {Schneider}, \citenamefont {Torgrimsson},\ and\
  \citenamefont {Schützhold}(2018)}]{Schneider18}%
  \BibitemOpen
  \bibfield  {author} {\bibinfo {author} {\bibfnamefont {C.}~\bibnamefont
  {Schneider}}, \bibinfo {author} {\bibfnamefont {G.}~\bibnamefont
  {Torgrimsson}},\ and\ \bibinfo {author} {\bibfnamefont {R.}~\bibnamefont
  {Schützhold}},\ }\href@noop {} {\bibfield  {journal} {\bibinfo  {journal}
  {\prd}\ }\textbf {\bibinfo {volume} {98}},\ \bibinfo {pages} {085009}
  (\bibinfo {year} {2018})}\BibitemShut {NoStop}%
\bibitem [{\citenamefont {Franchino-Viñas}\ and\ \citenamefont
  {Gies}(2019)}]{FranchinoVinas19}%
  \BibitemOpen
  \bibfield  {author} {\bibinfo {author} {\bibfnamefont {S.}~\bibnamefont
  {Franchino-Viñas}}\ and\ \bibinfo {author} {\bibfnamefont {H.}~\bibnamefont
  {Gies}},\ }\href@noop {} {\bibfield  {journal} {\bibinfo  {journal} {\prd}\
  }\textbf {\bibinfo {volume} {100}},\ \bibinfo {pages} {105020} (\bibinfo
  {year} {2019})}\BibitemShut {NoStop}%
\bibitem [{\citenamefont {Corradini}\ and\ \citenamefont
  {Muratori}(2020)}]{Corradini20}%
  \BibitemOpen
  \bibfield  {author} {\bibinfo {author} {\bibfnamefont {O.}~\bibnamefont
  {Corradini}}\ and\ \bibinfo {author} {\bibfnamefont {M.}~\bibnamefont
  {Muratori}},\ }\href@noop {} {\bibfield  {journal} {\bibinfo  {journal} {{J.\
  High Energy Phys.}}\ }\textbf {\bibinfo {volume} {2020}},\ \bibinfo {pages}
  {169} (\bibinfo {year} {2020})}\BibitemShut {NoStop}%
\bibitem [{\citenamefont {Rajeev}(2021)}]{Rajeev21}%
  \BibitemOpen
  \bibfield  {author} {\bibinfo {author} {\bibfnamefont {K.}~\bibnamefont
  {Rajeev}},\ }\href@noop {} {\bibfield  {journal} {\bibinfo  {journal} {\prd}\
  }\textbf {\bibinfo {volume} {104}},\ \bibinfo {pages} {105014} (\bibinfo
  {year} {2021})}\BibitemShut {NoStop}%
\bibitem [{\citenamefont {Bastianelli}\ and\ \citenamefont
  {Paciarini}(2024)}]{Bastianelli24}%
  \BibitemOpen
  \bibfield  {author} {\bibinfo {author} {\bibfnamefont {F.}~\bibnamefont
  {Bastianelli}}\ and\ \bibinfo {author} {\bibfnamefont {M.~D.}\ \bibnamefont
  {Paciarini}},\ }\href@noop {} {\bibfield  {journal} {\bibinfo  {journal}
  {{Class. Quantum Grav.}}\ }\textbf {\bibinfo {volume} {41}},\ \bibinfo
  {pages} {115002} (\bibinfo {year} {2024})}\BibitemShut {NoStop}%
\bibitem [{\citenamefont {Scardicchio}\ and\ \citenamefont
  {Jaffe}(2006)}]{Scardiccio2006}%
  \BibitemOpen
  \bibfield  {author} {\bibinfo {author} {\bibfnamefont {A.}~\bibnamefont
  {Scardicchio}}\ and\ \bibinfo {author} {\bibfnamefont {R.~L.}\ \bibnamefont
  {Jaffe}},\ }\href@noop {} {\bibfield  {journal} {\bibinfo  {journal} {Nucl.
  Phys. B}\ }\textbf {\bibinfo {volume} {743}},\ \bibinfo {pages} {249}
  (\bibinfo {year} {2006})}\BibitemShut {NoStop}%
\bibitem [{\citenamefont {Glasserman}(2004)}]{Glasserman2004}%
  \BibitemOpen
  \bibfield  {author} {\bibinfo {author} {\bibfnamefont {P.}~\bibnamefont
  {Glasserman}},\ }\href@noop {} {\emph {\bibinfo {title} {Monte Carlo Methods
  in Financial Engineering}}}\ (\bibinfo  {publisher} {Springer},\ \bibinfo
  {year} {2004})\BibitemShut {NoStop}%
\bibitem [{\citenamefont {Asmussen}\ and\ \citenamefont
  {Glynn}(2007)}]{Asmussen2007}%
  \BibitemOpen
  \bibfield  {author} {\bibinfo {author} {\bibfnamefont {S.}~\bibnamefont
  {Asmussen}}\ and\ \bibinfo {author} {\bibfnamefont {P.~W.}\ \bibnamefont
  {Glynn}},\ }\href@noop {} {\emph {\bibinfo {title} {Stochastic Simulation}}}\
  (\bibinfo  {publisher} {Springer},\ \bibinfo {year} {2007})\BibitemShut
  {NoStop}%
\bibitem [{\citenamefont {Matacz}(2002)}]{Matacz2000}%
  \BibitemOpen
  \bibfield  {author} {\bibinfo {author} {\bibfnamefont {A.}~\bibnamefont
  {Matacz}},\ }\href@noop {} {\bibfield  {journal} {\bibinfo  {journal} {{J.\
  Comput.\ Finance}}\ }\textbf {\bibinfo {volume} {6}},\ \bibinfo {pages} {79}
  (\bibinfo {year} {2002})}\BibitemShut {NoStop}%
\bibitem [{\citenamefont {Predescu}(2003)}]{Predescu03}%
  \BibitemOpen
  \bibfield  {author} {\bibinfo {author} {\bibfnamefont {C.}~\bibnamefont
  {Predescu}},\ }\href@noop {} {\bibfield  {journal} {\bibinfo  {journal} {{J.\
  Math.\ Phys.}}\ }\textbf {\bibinfo {volume} {44}},\ \bibinfo {pages} {1226}
  (\bibinfo {year} {2003})}\BibitemShut {NoStop}%
\bibitem [{\citenamefont {Feynman}(1948)}]{Feynman1948}%
  \BibitemOpen
  \bibfield  {author} {\bibinfo {author} {\bibfnamefont {R.~P.}\ \bibnamefont
  {Feynman}},\ }\href@noop {} {\bibfield  {journal} {\bibinfo  {journal} {Rev.
  Mod. Phys.}\ }\textbf {\bibinfo {volume} {20}},\ \bibinfo {pages} {367}
  (\bibinfo {year} {1948})}\BibitemShut {NoStop}%
\bibitem [{\citenamefont {Kac}(1949)}]{Kac1949}%
  \BibitemOpen
  \bibfield  {author} {\bibinfo {author} {\bibfnamefont {M.}~\bibnamefont
  {Kac}},\ }\href@noop {} {\bibfield  {journal} {\bibinfo  {journal} {Trans.
  Amer. Math. Soc.}\ }\textbf {\bibinfo {volume} {64}},\ \bibinfo {pages} {1}
  (\bibinfo {year} {1949})}\BibitemShut {NoStop}%
\bibitem [{\citenamefont {Durrett}(1996)}]{Durrett1996}%
  \BibitemOpen
  \bibfield  {author} {\bibinfo {author} {\bibfnamefont {R.}~\bibnamefont
  {Durrett}},\ }\href@noop {} {\emph {\bibinfo {title} {Stochastic Calculus: A
  Practial Introduction}}}\ (\bibinfo  {publisher} {CRC Press},\ \bibinfo
  {year} {1996})\BibitemShut {NoStop}%
\bibitem [{\citenamefont {Karatzas}\ and\ \citenamefont
  {Shreve}(1991)}]{Karatzas1991}%
  \BibitemOpen
  \bibfield  {author} {\bibinfo {author} {\bibfnamefont {I.}~\bibnamefont
  {Karatzas}}\ and\ \bibinfo {author} {\bibfnamefont {S.~E.}\ \bibnamefont
  {Shreve}},\ }\href@noop {} {\emph {\bibinfo {title} {Brownian Motion and
  Stochastic Calculus}}}\ (\bibinfo  {publisher} {Spring-Verlag},\ \bibinfo
  {year} {1991})\BibitemShut {NoStop}%
\bibitem [{\citenamefont {Babb}, \citenamefont {Klimchitskaya},\ and\
  \citenamefont {Mostepanenko}(2004)}]{Babb2004}%
  \BibitemOpen
  \bibfield  {author} {\bibinfo {author} {\bibfnamefont {J.~F.}\ \bibnamefont
  {Babb}}, \bibinfo {author} {\bibfnamefont {G.~L.}\ \bibnamefont
  {Klimchitskaya}},\ and\ \bibinfo {author} {\bibfnamefont {V.~M.}\
  \bibnamefont {Mostepanenko}},\ }\href@noop {} {\bibfield  {journal} {\bibinfo
   {journal} {Phys.\ Rev.\ A}\ }\textbf {\bibinfo {volume} {70}},\ \bibinfo
  {pages} {042901} (\bibinfo {year} {2004})}\BibitemShut {NoStop}%
\bibitem [{\citenamefont {Schoger}\ \emph {et~al.}(2022)\citenamefont
  {Schoger}, \citenamefont {Spreng}, \citenamefont {Ingold}, \citenamefont
  {Lambrecht}, \citenamefont {Neto},\ and\ \citenamefont
  {Reynaud}}]{Schoger2022}%
  \BibitemOpen
  \bibfield  {author} {\bibinfo {author} {\bibfnamefont {T.}~\bibnamefont
  {Schoger}}, \bibinfo {author} {\bibfnamefont {B.}~\bibnamefont {Spreng}},
  \bibinfo {author} {\bibfnamefont {G.-L.}\ \bibnamefont {Ingold}}, \bibinfo
  {author} {\bibfnamefont {A.}~\bibnamefont {Lambrecht}}, \bibinfo {author}
  {\bibfnamefont {P.~A.~M.}\ \bibnamefont {Neto}},\ and\ \bibinfo {author}
  {\bibfnamefont {S.}~\bibnamefont {Reynaud}},\ }\href@noop {} {\bibfield
  {journal} {\bibinfo  {journal} {Int.\ J.\ Mod.\ Phys.\ A}\ }\textbf {\bibinfo
  {volume} {37}},\ \bibinfo {pages} {2241005} (\bibinfo {year}
  {2022})}\BibitemShut {NoStop}%
\bibitem [{\citenamefont {Borodin}\ and\ \citenamefont
  {Salminen}(2002)}]{Borodin2002}%
  \BibitemOpen
  \bibfield  {author} {\bibinfo {author} {\bibfnamefont {A.~N.}\ \bibnamefont
  {Borodin}}\ and\ \bibinfo {author} {\bibfnamefont {P.}~\bibnamefont
  {Salminen}},\ }\href@noop {} {\emph {\bibinfo {title} {Handbook of Brownian
  Motion---Facts and Formulae}}},\ \bibinfo {edition} {2nd}\ ed.\ (\bibinfo
  {publisher} {Birkh\"auser},\ \bibinfo {year} {2002})\BibitemShut {NoStop}%
\end{thebibliography}

\end{document}